\documentclass{article}

\usepackage{microtype}
\usepackage{graphicx}
\usepackage{subcaption}
\usepackage{booktabs} 
\usepackage{url}            
\usepackage{hyperref}
\usepackage{amsfonts}       
\usepackage{nicefrac}       
\usepackage{microtype}      
\usepackage{xcolor}         
\usepackage{multirow}
\usepackage[ruled,vlined]{algorithm2e} 

\usepackage{enumitem}
\usepackage[preprint]{icml2026}

\usepackage{amsmath}
\usepackage{amssymb}
\usepackage{mathtools}
\usepackage{amsthm}
\usepackage{wrapfig} 

\usepackage[capitalize,noabbrev]{cleveref}

\theoremstyle{plain}

\theoremstyle{definition}

\theoremstyle{remark}

\usepackage[textsize=tiny]{todonotes}

\icmltitlerunning{Sample-Efficient Diffusion-based Control of Complex Physics Systems}

\begin{document}

\twocolumn[
  \icmltitle{Sample-Efficient Diffusion-based Control of Complex Physics Systems}



  \icmlsetsymbol{equal}{*}

  \begin{icmlauthorlist}
    \icmlauthor{Hongyi Chen}{xxx,comp}
    \icmlauthor{Jingtao Ding}{yyy}
    \icmlauthor{Jianhai Shu}{yyy}
    \icmlauthor{Xinchun Yu}{xxx}
    \icmlauthor{Xiaojun Liang}{comp}
    \icmlauthor{Yong Li}{yyy}
    \icmlauthor{Xiao-Ping Zhang}{xxx}

  \end{icmlauthorlist}
  \icmlaffiliation{xxx}{Shenzhen International Graduate School, Tsinghua University, Shenzhen, China}
  \icmlaffiliation{yyy}{Department of EE, Beijing, China}
  \icmlaffiliation{comp}{Pengcheng Laboratory, Shenzhen, China}

  \icmlcorrespondingauthor{Jingtao Ding}{dingjt15@tsinghua.org.cn}

  \icmlkeywords{Machine Learning, ICML}

  \vskip 0.3in
]



\printAffiliationsAndNotice{}  

\begin{abstract}
Controlling complex physics systems is important in diverse domains. While diffusion-based methods have demonstrated advantages over classical model-based approaches and myopic sequential learning methods in achieving global trajectory consistency, they are limited by sample efficiency.
This paper presents SEDC (Sample-Efficient Diffusion-based Control), a novel framework addressing core challenges in complex physics systems: high-dimensional state-control spaces, strong nonlinearities, and the gap between non-optimal training data and near-optimal control laws.
Our approach introduces a novel control paradigm by architecturally decoupling state-control modeling and decomposing dynamics, while a guided self-finetuning process iteratively refines the control law towards optimality.  
We validate SEDC across diverse complex nonlinear systems, including high-dimensional fluid dynamics (Burgers), chaotic synchronization networks (Kuramoto), and real-world power grid stability control (Swing Equation).
Our method achieves 39.5\%-47.3\% better control accuracy than state-of-the-art baselines while using only 10\% of the training samples. The implementation is available at \href{https://anonymous.4open.science/r/DIFOCON-C019}{here}.
\end{abstract}

\section{Introduction}
\label{Introduction}
The control of complex physics systems plays a critical role across diverse domains, from industrial automation~\citep{baggio2021data} to biological networks~\citep{gu2015controllability}. Given the challenges in deriving governing equations for empirical systems, data-driven control methods—which design control modules directly based on experimental data collected from the system, bypassing the need for explicit mathematical modeling—have gained prominence for their robust real-world applicability~\citep{baggio2021data,janner2022planning,ajay2022conditional,zhou2024adaptive,liang2023adaptdiffuser,wei2024generative}.

Traditional Proportional-Integral-Derivative (PID)~\citep{li2006pid} and Model Predictive Control (MPC)~\citep{schwenzer2021review} methods are limited in complex nonlinear systems. PID controllers struggle with nonlinearities, while MPC's performance depends on model accuracy and is computationally intensive for long-horizon tasks. Data-driven approaches have emerged to address these issues, including supervised learning, sequential learning approaches, and diffusion-based methods. Supervised trajectory fitting baselines (e.g., BC~\citep{pomerleau1988alvinn}) and sequential learning methods (e.g., BPPO~\citep{zhuang2023behavior}) often make myopic, step-by-step decisions, leading to suboptimal outcomes in long-horizon tasks. In contrast, diffusion-based methods~\citep{janner2022planning,ajay2022conditional,zhou2024adaptive,liang2023adaptdiffuser,wei2024generative} treat control as a \textit{global trajectory generation} problem. By generating the entire control plan in a single sample, they achieve comprehensive optimization over the full trajectory, avoiding the pitfalls of iterative methods and enabling superior long-term performance.

The success of diffusion models is dependent on their ability to learn complex trajectory distributions. However, in practice, the available trajectory data is often non-optimal and sparse due to collection under empirical rules or random actuation and high operational costs. This presents a key challenge for diffusion-based methods: synthesizing effective control laws from limited and suboptimal data.
\textbf{First, limited data volume impedes sample-efficient modeling in high-dimensional physics systems.} Existing diffusion-based controllers \citep{wei2024cl,wei2024generative,hu2025uncertain} attempt to directly generate long-term~($T$ steps) state-control trajectories by learning a $T\times(P+M)$-dimensional distribution of system states $y^P$ and control inputs $u^M$.  This joint distribution implicitly encodes system dynamics of state transitions under external control inputs, which often leads to physically inconsistent trajectories when training samples are insufficient~\cite{janner2022planning}.
\textbf{Second, synthesizing control laws for nonlinear systems remains an open challenge both theoretically and practically.} 
Traditional analytical methods~\citep{baggio2021data} designed for linear systems fail to perform robustly when applied to nonlinear systems.
While diffusion-based approaches~\citep{janner2022planning,ajay2022conditional,zhou2024adaptive,zhong2025constrained,hu2025uncertain} employ deep neural networks (e.g., U-Net architectures) as denoising modules to capture nonlinearity, synthesizing effective control laws from limited data remains particularly challenging for complex systems with strong nonlinearity, such as fluid dynamics~\cite{brunton2020machine} and power grids~\cite{baggio2021data}.
\textbf{Third, extracting improved control laws from non-optimal training data poses fundamental difficulties.} Diffusion-based methods~\citep{janner2022planning} struggle when training data significantly deviates from optimal solutions. Although recent work~\citep{wei2024generative} introduces reweighting mechanism to expand the solution space during generation, discovering truly near-optimal control laws remains elusive without explicit optimization guidance.

To address these challenges, we propose SEDC (\textbf{S}ample-\textbf{E}fficient \textbf{D}iffusion-based \textbf{C}ontrol), a novel diffusion-based framework for synthesizing control laws of complex physics systems with limited, non-optimal data. At its core, SEDC reformulates the control problem as a denoising diffusion process that samples control sequences optimized for reaching desired states while minimizing energy consumption. We then solve the sample efficiency challenge by addressing its three key aspects.
To address the curse of dimensionality, we introduce Decoupled State Diffusion (DSD), which simplifies the modeling complexity of the generative task. By diffusing only on the more structured state space, rather than the complex joint state-control space, DSD achieves higher sample efficiency. A separate inverse dynamics model is then used to ensure physical consistency.
To tackle strong nonlinearity, we propose Dual-Mode Decomposition (DMD) by designing a dual-UNet denoising module with residual connections. This architecture decomposes system dynamics into hierarchical linear and nonlinear components, enabling structured modeling of complex systems.
To bridge the gap between non-optimal training data and optimal control laws, we introduce the Guided Self-finetuning (GSF) mechanism, which progressively synthesizes guided control trajectories for iterative finetuning, facilitating manifold expansion beyond initial training data and convergence toward near-optimal control laws.

Our contributions are summarized as follows:
\begin{itemize}[leftmargin=*, noitemsep, topsep=0pt]
    \item We introduce a data-driven framework that significantly enhances sample efficiency in controlling high-dimensional nonlinear systems via diffusion models.
    \item Experiments on three complex nonlinear systems show that SEDC outperforms traditional, sequential, and diffusion-based baselines, achieving a 39.5\%–47.3\% improvement in accuracy with a better accuracy-energy balance.
    \item SEDC matches top-tier performance using only 10\% of training data. We further demonstrate its scalability on 2D PDEs and robustness to non-invertible systems.
    \item We confirm the method's computational efficiency and provide extensive ablation studies to verify the specific contribution of each architectural component.
\end{itemize}

\begin{figure*}[t]
    \centering
    \includegraphics[width=0.92\linewidth]{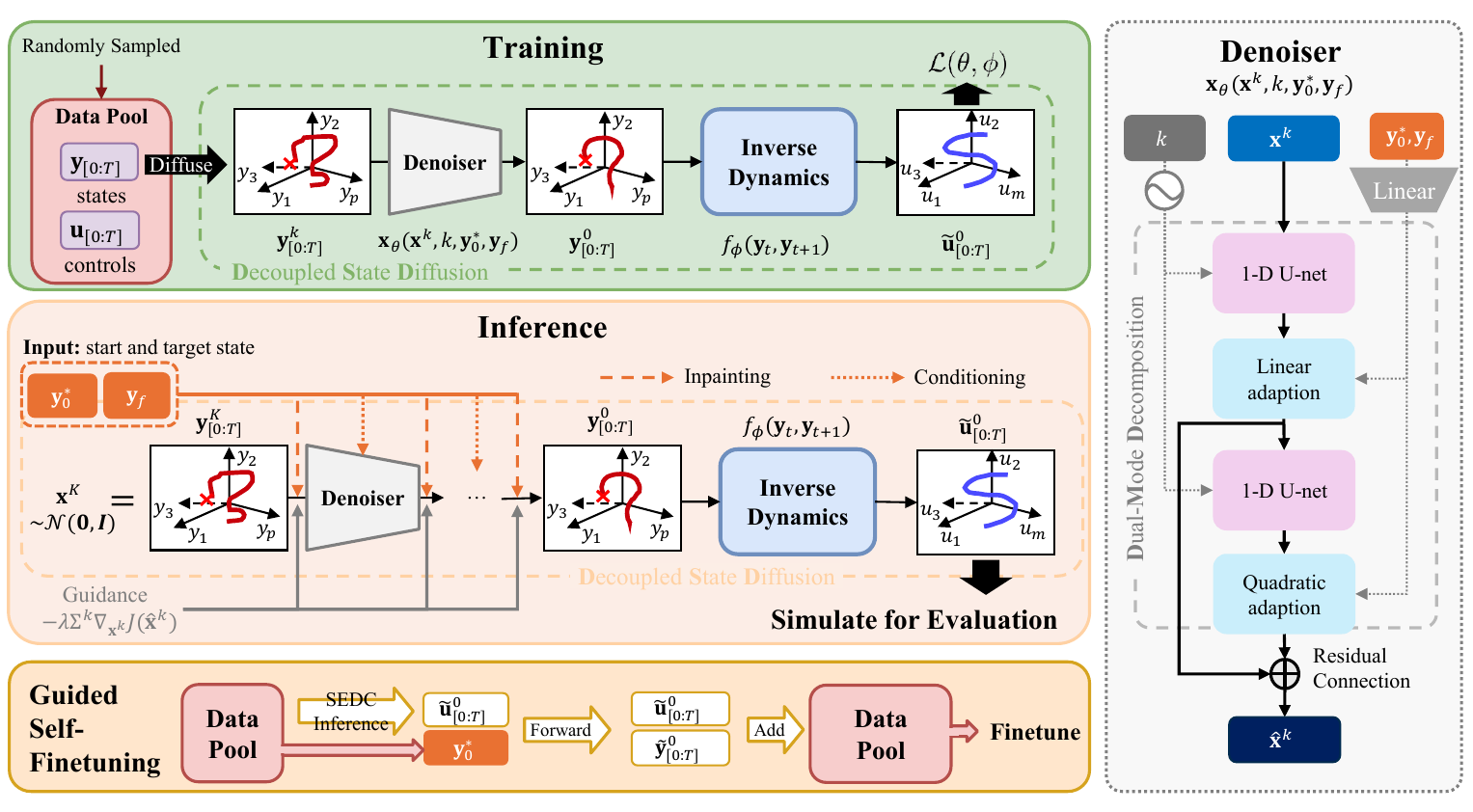}    
    \caption{Overview of SEDC. The framework consists of a training/finetuning (top panel), inference process (middle panel) and finetuning process (bottom panel). The core denoising network employs our Dual-Mode Decomposition (DMD) architecture (right panel). Both training and inference leverage Decoupled State Diffusion (DSD) by diffusing only on states and using a separate inverse dynamics model to recover controls.}
    \label{fig:model}
    \vspace{-10pt}
\end{figure*}

\section{Related Work}
\label{rw}

Data-driven control encompasses various paradigms, which can be broadly categorized by their approach to trajectory generation.
One major paradigm is \textit{iterative, feedback-based control}. Classical methods like PID controllers \citep{li2006pid} operate via real-time error correction but face limitations in high-dimensional complex scenarios. More contemporary approaches center on system identification, such as Dynamic Mode Decomposition \citep{tu2013dynamic} and Koopman operator theory \citep{mauroy2020koopman}. These methods first learn an explicit dynamics model from data and then design a controller, often within a Model Predictive Control (MPC) framework \citep{schwenzer2021review}. While robust for real-time adaptation, this two-stage paradigm is susceptible to compounding errors, particularly in sample-scarce settings. Inaccuracies in the learned model can accumulate over long horizons, degrading control performance. Sequential learning approaches such as supervised trajectory fitting \citep{pomerleau1988alvinn} and reinforcement learning \citep{haarnoja2018soft,zhuang2023behavior} offer adaptive methods but can struggle with long-horizon credit assignment and compounding errors due to their myopic, step-by-step nature~\cite{ross2011reduction}.

In contrast, \textit{global trajectory planning} reframes control as a holistic generation problem, for which denoising diffusion models \citep{ho2020denoising,dhariwal2021diffusion,kong2020diffwave,ho2022video} have emerged as a powerful tool. By generating the entire control plan as a single, coherent sample, these methods capture long-term dependencies and avoid the pitfalls of iterative error accumulation. Seminal works \citep{janner2022planning, ajay2022conditional} demonstrated this potential in robotics, but their generic architectures struggle with strong nonlinearities; our Dual-Mode Decomposition (DMD) architecture addresses this with a structured inductive bias. Subsequent research has tackled specific limitations. DiffPhyCon \citep{wei2024generative} uses reweighting to synthesize trajectories extending beyond the training distribution. This approach, however, requires training separate denoising networks to model decomposed energy functions (one for the prior and one for the conditional distribution). Moreover, its joint state-control modeling can exacerbate the curse of dimensionality, which our Decoupled State Diffusion (DSD) alleviates by diffusing over the state space alone. To bridge the data-optimality gap, AdaptDiffuser \citep{liang2023adaptdiffuser} fine-tunes on discriminator-filtered trajectories, whereas our Guided Self-finetuning (GSF) employs a simpler, filter-free loop. Beyond standard architectures, advanced approaches have emerged to address specific physical challenges—such as WDNO \citep{hu2024wavelet} for multi-resolution dynamics and SafeDiffCon \citep{hu2025uncertain} for safety constraints. However, these methods typically prioritize simulation fidelity or boundary satisfaction, rather than the sample-efficient synthesis of near-optimal control policies from sparse, suboptimal data. 

\section{Backgrounds}

\subsection{Problem Setting}

The dynamics of a controlled complex system can be represented by the differential equation $\dot{\mathbf{y}}_t = \Phi(\mathbf{y}_t,\mathbf{u}_t)$, where $\mathbf{y}_t \in \mathbb{R}^N$ represents the system state and $\mathbf{u}_t \in \mathbb{R}^M$ denotes the control input.
We assume the system satisfies the controllability condition without loss of generality: for any initial state $\mathbf{y}_0^*$ and target state $\mathbf{y}_f$, there exists a finite time $T$ and a corresponding control input $\mathbf{u}$ that can drive the system from $\mathbf{y}_0^*$ to $\mathbf{y}_f$. This assumption ensures the technical feasibility of our control objectives.
In practical applications, beyond achieving state transitions, we need to optimize the energy consumption during the control process. The energy cost can be quantified using the L2-norm integral of the control input: $J(\mathbf{y},\mathbf{u}) = \int_0^T |\mathbf{u}(t')|^2 dt'$.
Consider a dataset $D = \{\mathbf{u}^{(i)}, \mathbf{y}^{(i)}\}_{i=1}^P$ containing $P$ non-optimal control trajectories, where each trajectory consists of: (1) complete state trajectories $\mathbf{y}^{(i)}={\mathbf{y}^{(i)}_0,...,\mathbf{y}^{(i)}_T}$ sampled at fixed time intervals; (2) corresponding control input sequences $\mathbf{u}^{(i)}={\mathbf{u}^{(i)}_0,...,\mathbf{u}^{(i)}_{  T-1}}$.
Our objective is satisfying the boundary constraint (reaching $\mathbf{y}_f$ from $\mathbf{y}_0^*$) and minimize the energy cost $J(\mathbf{y},\mathbf{u})$ within the feasible solution space. Formally, this is a constrained optimization problem:
\begin{equation}
\begin{aligned}
\min_\mathbf{u} \quad & J(\mathbf{y},\mathbf{u}) \\
\text{s.t.} \quad & \mathbf{y}_T = \mathbf{y}_f \quad \text{(Target Satisfaction)}, \\
& \Psi(\mathbf{u},\mathbf{y}) = 0 \quad \text{(Dynamics Consistency)},
\end{aligned}
\end{equation}
where $\mathbf{y}\in \mathbb{R}^{T\times N}$ is the corresponding complete state trajectory given $\mathbf{y}_0=\mathbf{y}_0^*$. $\Psi(\mathbf{u},\mathbf{y}) = 0$ represents the system dynamics constraint implicitly defined by dataset $D$. This constraint effectively serves as a data-driven representation of the unknown dynamics equation $\dot{\mathbf{y}}_t = \Phi(\mathbf{y}_t,\mathbf{u}_t)$.

Our key idea is to train a diffusion-based model to directly produce near-optimal control trajectories $\mathbf{u}_{[0:T-1]}$, providing a starting state $\mathbf{y}_0^*$, the target $\mathbf{y}_f$ and optimized by the cost $J$. Next, we summarize the details of the diffusion-based framework.

\subsection{Diffusion Model}

Diffusion models have become the leading generative models, showing exceptional results across image synthesis, audio generation and other applications \citep{ho2020denoising,dhariwal2021diffusion,song2019generative}. These models operate by progressively adding noise to sequential data in the forward process and then learning to reverse this noise corruption through a denoising process.
We formulate control as a conditional generative task. The \textit{forward process} progressively corrupts a clean trajectory $\mathbf{x}^0$ into Gaussian noise $\mathbf{x}^K \sim \mathcal{N}(\mathbf{0}, \mathbf{I})$ over $K$ steps.
The \textit{reverse process} learns a denoising network $\mathbf{x}_\theta$ to reconstruct the clean trajectory from noise, conditioned on system constraints (e.g., $\mathbf{y}_0^*, \mathbf{y}_f$).
Following standard practices \citep{janner2022planning}, we train the network to directly predict the clean trajectory $\hat{\mathbf{x}}^0$ at each step $k$ by minimizing the simplified objective:$
    \mathcal{L}(\theta) = \mathbb{E}_{k, \mathbf{x}^0, \mathbf{\epsilon}} \left[ || \mathbf{x}^0 - \mathbf{x}_\theta(\mathbf{x}^k, k, \mathbf{y}_0^*, \mathbf{y}_f) ||^2 \right],$
where $k \sim \mathcal{U}\{1, \dots, K\}$ is the diffusion step, and $\mathbf{x}^k = \sqrt{\bar{\alpha}^k}\mathbf{x}^0 + \sqrt{1-\bar{\alpha}^k}\mathbf{\epsilon}$ is the noisy input constructed from variance schedule $\bar{\alpha}^k$.

\section{SEDC: the Proposed Method}

\subsection{Controlling with Diffusion Models}

As shown in Figure~\ref{fig:model}, SEDC reformulates the control problem as a conditional trajectory generation task. The core idea is to train a diffusion model, conditioned on the initial $\mathbf{y}_0^*$ and target $\mathbf{y}_f$ states, to directly generate a complete state trajectory $\mathbf{y}_{[0:T]}$. Subsequently, a corresponding control sequence $\mathbf{u}_{[0:T-1]}$ is derived from this state trajectory via a learned inverse dynamics model.

\textbf{Decoupled State Diffusion (DSD).} Jointly modeling the state-control distribution is highly sample-intensive and risks generating physically inconsistent trajectories. To address this, we propose Decoupled State Diffusion (DSD). As shown in Figure~\ref{fig:model}, we confine the diffusion process to the state trajectory $\mathbf{y}$ alone (i.e., $\mathbf{x}:=\mathbf{y}_{[0:T]}$), as state evolution is generally smoother and more structured. The corresponding control sequence is then derived from the generated state transitions via a separately trained inverse dynamics model, $f_\phi$:
$
\tilde{\mathbf{u}}_{t,\text{update}}^0 = f_\phi(\mathbf{y}_t^0, \mathbf{y}_{t+1}^0),
$
where the superscript $0$ denotes the final denoised output of the diffusion model. 

A common concern with deterministic inverse dynamics models $f_\phi$ is the potential for averaging valid controls in multi-modal scenarios (where distinct actions yield identical transitions), leading to physically invalid actions. However, Unlike joint modeling state and control, which risks generating out-of-distribution pairs, DSD ensures that the input to the inverse model $(\mathbf{y}_t, \mathbf{y}_{t+1})$ is always drawn from a valid, coherent manifold generated by the diffusion model~\cite{chen2025learning}. Even in settings with control non-uniqueness, the inverse model is only tasked with finding a feasible control that satisfies this valid transition, a significantly well-posed problem compared to unconstrained joint generation. We also evaluate and discuss SEDC's robustness to non-invertibility in Section~\ref{sec:non_invertible}.

We then optimize $f_\phi$ simultaneously with the denoiser. The loss function is:
\vspace{-3pt}
\begin{equation}
\begin{aligned}
\mathcal{L}(\theta, \phi) &:= \mathbb{E}_{\mathbf{x},k,\mathbf{y}_0^*,\mathbf{y}_f,\mathbf{\epsilon}}[||\boldsymbol{\mathbf{x}}-\boldsymbol{\mathbf{x}}_\theta(\boldsymbol{\mathbf{x}}^k, k,\mathbf{y}_0^*,\mathbf{y}_f)||^2]\\&+\mathbb{E}_{\mathbf{y}_t,\mathbf{u}_t,\mathbf{y}_{t+1}}[||\mathbf{u}_t-f_\phi(\mathbf{y}_t, \mathbf{y}_{t+1})||^2],
\end{aligned}
\end{equation}
where $\mathbf{y}_t,\mathbf{u}_t,\mathbf{y}_{t+1}$ are sampled from the dataset. After training, the denoising process generates a state trajectory $\mathbf{y}_{0:T}$ that is both physically plausible and conditioned on the start and target states. To further refine this process, we introduce two key mechanisms: inpainting for hard constraint satisfaction and gradient guidance for soft optimization.

\textbf{Target-Conditioning via Inpainting.} To ensure the generated trajectory strictly adheres to the given initial state $\mathbf{y}_0^*$ and target state $\mathbf{y}_f$, we treat the problem as a form of trajectory inpainting. While these states are provided as conditions to the denoising network, we enforce them as hard constraints during the sampling process. Specifically, at each denoising step $k$, after sampling a potential trajectory $\mathbf{x}^{k-1} \sim p_\theta(\mathbf{x}^{k-1} | \mathbf{x}^k, \mathbf{y}_0^*, \mathbf{y}_f)$, we replace its start and end points with the ground truth values: $\mathbf{x}^{k-1}_0 \leftarrow \mathbf{y}_0^*$ and $\mathbf{x}^{k-1}_T \leftarrow \mathbf{y}_f$. This technique, analogous to inpainting in image generation \citep{lugmayr2022repaint}, guarantees that the final output satisfies the boundary conditions.

\textbf{Cost Optimization via Gradient Guidance.} Beyond satisfying the boundary constraints, we aim to find a trajectory that minimizes a given cost function $J$ (e.g., control energy). We achieve this through inference-time gradient guidance. Building upon the inpainting-enforced sampling, we further modify the mean of the denoising distribution by incorporating the cost gradient:
\vspace{-3pt}
\begin{equation}
\begin{aligned}
    \mathbf{\mu}_\theta(\mathbf{x}^k, k, \mathbf{y}_0^*, \mathbf{y}_f) =& \frac{\sqrt{\bar{\alpha}^{k-1}}\beta^k}{1-\bar{\alpha}^k}\hat{\mathbf{x}}^k + \frac{\sqrt{\alpha^k}(1-\bar{\alpha}^{k-1})}{{1-\bar{\alpha}^k}}\mathbf{x}^k \\& - \lambda\Sigma^k\nabla_{\mathbf{x}^k}J(\hat{\mathbf{x}}^k(\mathbf{x}^k)),
\end{aligned}
\end{equation}
where $\lambda$ controls guidance strength and the superscript $k$ of $\hat{\mathbf{x}}^k:=\hat{\mathbf{y}}_{[0:T]}^k$ denotes the clean output from the denoiser at step $k$. Since our diffusion model operates on states only, we recover the control sequence $\tilde{\mathbf{u}}_t^k = f_\phi(\hat{\mathbf{y}}_t^k, \hat{\mathbf{y}}_{t+1}^k)$ to compute the cost $J({\mathbf{u}})$ at each step. This combined approach of inpainting and guidance ensures the final generated trajectory is not only feasible and satisfies hard constraints but is also optimized for the desired cost objective.

\subsection{Dual-Mode Decomposition~(DMD) for Denoiser}

In this section, we propose our denoising network design, DMD, that decomposes the modeling of linear and nonlinear modes in the sampled trajectory by a dual-Unet architecture, as shown in Figure~\ref{fig:model}. Our design draws inspiration from control theory. For linear systems, \citet{yan2012controlling} demonstrated that optimal control signals have a linear relationship with a specific linear combination $\mathbf{y}_c$ of initial and target states. Building upon this insight, we develop a framework where a bias-free linear layer first learns this crucial linear combination $\mathbf{y}_c$ from the initial state $\mathbf{y}_0$ and target state $\mathbf{y}_f$.
Then, our module decomposes the prediction of the clean sampled trajectory into linear and nonlinear modes, overcoming the limitations of single-network approaches that struggle to model both simultaneously. The theoretical foundation is as follows: $\mathbf{y}_c$ is the conditional input, and our denoiser is designed to output the clean state trajectory $\hat{\mathbf{x}}^0$, expressed as a vector function $\mathbf{f}(\mathbf{y}_c)$. It admits a vector Taylor expansion at $\mathbf{y}_c=\mathbf{0}$ as $\hat{\mathbf{x}}^0 = \mathbf{f}(\mathbf{y}_c) = \mathbf{C}_1 \mathbf{y}_c + \mathbf{y}_c^T \mathbf{C}_2 \mathbf{y}_c + \mathcal{O}(||\mathbf{y}_c||^3)$. For linear systems, only the first-order term remains. For nonlinear systems, by neglecting higher-order terms for simplicity, we can decompose the prediction into linear and nonlinear quadratic modes.

This decomposition imposes a strong inductive bias: modeling a dominant linear part and a subtle nonlinear correction is a more stable and sample-efficient task than forcing a monolithic network to learn the entire complex function from scratch. We implement this with a dual-UNet architecture, as illustrated in Figure~\ref{fig:model}. The network is conditioned on a vector $\mathbf{y}_c = f_c([\mathbf{y}_0, \mathbf{y}_f]) \in \mathbb{R}^{B\times C_1}$, which is generated from the start and target states via a bias-less linear layer $f_c$.

Given the noisy trajectory $\mathbf{x}^k \in \mathbb{R}^{B\times T\times N}$ and time embedding $\mathbf{k}_{\text{emb}}$, the first stage produces a linear prediction $\mathbf{O}_1 \in \mathbb{R}^{B\times T\times N}$ from first-order coefficients $\mathbf{C}_1$:
\begin{equation}
    \mathbf{C}_1 = \text{UNet}_1(\mathbf{x}^k,\mathbf{k}_{\text{emb}}),
    \quad
    \mathbf{O}_1 = \text{reshape}(\mathbf{C}_1)\cdot \mathbf{y}_c.
\end{equation}
In the second stage, UNet$_2$ combines the input $\mathbf{x}^k$ and intermediate features $\mathbf{C}_1$ to generate quadratic coefficients $\mathbf{C}_2$, which yield a nonlinear correction term $\mathbf{O}_2 \in \mathbb{R}^{B\times T\times N}$:
\begin{equation}
    \mathbf{C}_2 = \text{UNet}_2([\mathbf{x}^k, \mathbf{C}_1],\mathbf{k}_{\text{emb}}),
    \quad
    \mathbf{O}_2 = \mathbf{y}_c^T\cdot\text{reshape}(\mathbf{C}_2)\cdot \mathbf{y}_c.
\end{equation}
The final denoised prediction is the sum of these components:
\begin{equation}
    \hat{\mathbf{x}}^0 = \mathbf{O}_1+\mathbf{O}_2 \in \mathbb{R}^{B\times T\times N}.
\end{equation}
Here, $\mathbf{C}_1 \in \mathbb{R}^{B\times T\times (N\times C_1)}$ and $\mathbf{C}_2 \in \mathbb{R}^{B\times T\times (C_1\times N \times C_1)}$ represent the learned coefficient tensors.

\subsection{Guided Self-finetuning~(GSF)}

Randomly generated training data cannot guarantee coverage of optimal scenarios. To generate near-optimal controls that may deviate significantly from the training distribution, we propose leveraging the model's initially generated data (under the guidance of the cost function), which naturally deviates from the training distribution toward optimality, for iterative retraining to systematically expand the manifold coverage. 
This approach maintains physical consistency by ensuring generated samples adhere to the underlying system dynamics.

As shown in Figure~\ref{fig:model}, our methodology involves extracting control sequences from the generated samples (i.e., the output of inverse dynamics $\tilde{\mathbf{u}}^0$) and reintroducing them into the ground-truth simulator (or the physical environment if accessible) to generate corresponding state sequences $\tilde{\mathbf{y}}^0$. While this introduces an online interaction component, it is highly practical in scientific control tasks (e.g., fluid dynamics) where forward simulation is possible but computationally expensive to run exhaustively. GSF serves as a sample-efficient bridge between offline initialization and optimal control. Together, we add the renewed $[\tilde{\mathbf{u}}^0,\tilde{\mathbf{y}}^0]$ to the retrain data pool used for a new round of fine-tuning, notably without requiring explicit system parameter identification. We iterate this process over multiple rounds specified by a hyperparameter, systematically expanding the model's manifold support to progressively approach the optimal control law. Denote the sampling process under cost $J$'s guidance and the following interacting process as $[\tilde{\mathbf{u}}^0,\tilde{\mathbf{y}}^0]=\mathcal{S}(\mathbf{x}^K,\mathbf{y}_0^*,\mathbf{y}_f,J,\Phi)$. The process can be formulated as:
\begin{equation}
[\tilde{\mathbf{u}}^0,\tilde{\mathbf{y}}^0]=\mathcal{S}_{(\mathbf{x}^K,\mathbf{y}_0^*,\mathbf{y}_f)\sim D}(\mathbf{x}^K,\mathbf{y}_0^*,\mathbf{y}_f,J,\Phi),
\end{equation}
\begin{equation}
D = [D, [\tilde{\mathbf{u}}^0,\tilde{\mathbf{y}}^0]],
\end{equation}
where $D$ is the training set and $\mathcal{S}(\mathbf{x}^K,\mathbf{y}_0^*,\mathbf{y}_f,J,\Phi)$ denotes the full process of generating a guided trajectory using cost $J$ and then interacting with the system dynamics $\Phi$ to get the corresponding state sequence. We provide the algorithm form of SEDC in Appendix~\ref{sup:alg}.

\begin{figure*}[t]
    \centering
    \includegraphics[width=0.95\linewidth]{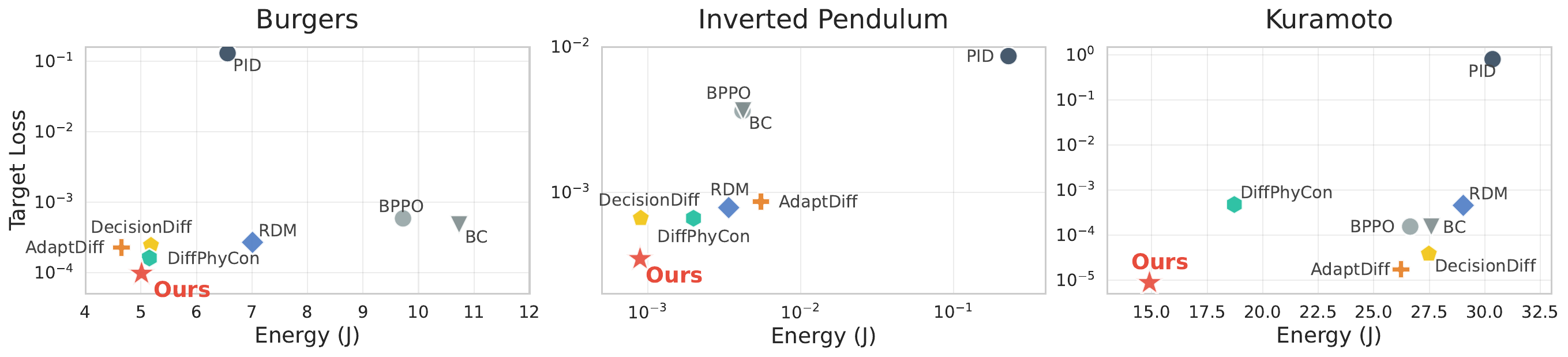}    
    \vspace{-5pt}
    \caption{Comparison of target loss and energy cost $J$ across different datasets. The closer the data point is to the bottom left, the better the performance.}
    \label{fig:overall}
\vspace{-1em}
\end{figure*}

\begin{figure*}[t]
    \centering
    \includegraphics[width=0.95\linewidth]{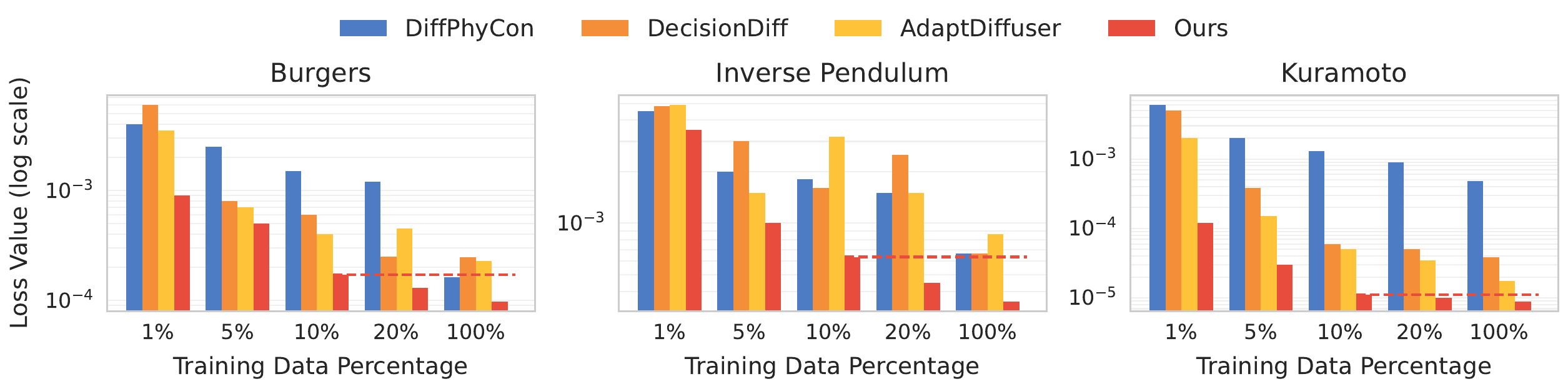}
    \vspace{-5pt}
    \caption{Sample-efficiency comparison on Burgers, Kuramoto and Inverse Pendulum dynamics.}
    \label{fig:efficiency}
\vspace{-1.5em}
    
\end{figure*}

\section{Experiments}

\subsection{Experiment Settings} 

We conducted experiments on multiple nonlinear physics systems, following the instructions in the previous works for data synthesis. These systems include: the 1-D Burgers dynamics~\citep{hwang2022solving,wei2024generative}, which is a high-dimensionality (128-dim, significantly larger than standard low-dimensional robotics benchmarks) dynamic system for studying nonlinear wave propagation and turbulent fluid flow; the Kuramoto dynamics~\citep{acebron2005kuramoto,baggio2021data,gupta2022learning}, which is essential for understanding synchronization phenomena in complex networks and coupled oscillator systems; the inverted pendulum dynamics~\citep{boubaker2013inverted}, a classic nonlinear control baseline. To further assess the scalability of our approach in extreme high-dimensional settings, we also evaluate SEDC on Jellyfish locomotion, a challenging 2D PDE control task involving complex fluid-solid interactions (detailed in Appendix~\ref{app:jellyfish}). For each system, we generated control/state trajectory data using the finite difference method and selected 50 trajectories as the test set. We assume full state observability throughout the experiments
. Detailed descriptions of the system dynamics equations and data synthesis procedures are provided in Appendix~\ref{app:b}. Implementation details and time consumption reports are provided in Appendix \ref{sup:imp}. 

We evaluate two metrics which is crucial in complex system control: \textbf{Target Loss}, the mean-squared-error~(MSE) of $\mathbf{y}_T$ and desired target $\mathbf{y}_f$, i.e. $\frac{1}{N}\|\mathbf{y}_T - \mathbf{y}_f\|^2$. (Note that $\mathbf{y}_T$ is obtained by simulating the real system using the control inputs generated by each method, along with the given initial state conditions, rather than extracted from the sample trajectories of the diffusion-based methods); \textbf{Energy} $J = \int_0^T |\mathbf{u}(t')|^2 dt'$, which measures the cumulative control effort required to achieve the target state. Lower values of both metrics indicate better performance.

\noindent\textbf{Baselines.} We evaluate SEDC against two categories of control methods.
\textit{Iterative Methods:} we include the classical \textbf{PID} controller \citep{li2006pid} and data-driven sequential learning approaches, including Behavioral Cloning (BC) \citep{pomerleau1988alvinn} and Batch Proximal Policy Optimization (BPPO) \citep{zhuang2023behavior}. 
\textit{Global Methods:} we also compare against state-of-the-art diffusion-based methods that generate holistic plans: DecisionDiffuser (DecisionDiff) \citep{ajay2022conditional}, AdaptDiffuser (AdaptDiff) \citep{liang2023adaptdiffuser}, RDM \citep{zhou2024adaptive}, and DiffPhyCon \citep{wei2024generative}.  These global methods avoid error accumulation by solving for the entire trajectory holistically. Detailed descriptions are included in Appendix~\ref{sup:baselines}.

\subsection{Overall Control Performance}

In Figure~\ref{fig:overall}, we compare different methods' performance across three dynamical systems using two-dimensional coordinate plots, where proximity to the lower-left corner indicates better trade-offs between control accuracy and energy efficiency. Since unstable control can lead to system failure regardless of energy efficiency, we prioritize control accuracy and report metrics at each method's minimum Target Loss.

\textbf{SEDC consistently achieves state-of-the-art performance.} Our method secures the position closest to the origin in all three datasets, demonstrating the best balance between accuracy and efficiency. Specifically, SEDC achieves the lowest Target Loss across all systems, outperforming the strongest baselines by 39.5\%, 49.4\%, and 47.3\% in the Burgers, Kuramoto, and IP systems, respectively. This highlights its superior capability in learning complex dynamics. In terms of energy cost, SEDC leads on the Kuramoto and IP systems and remains highly competitive on the Burgers system.

\textbf{Analysis of Baselines.} The results reveal clear performance tiers among different method families. Traditional PID control shows the poorest performance, as system complexity exacerbates the difficulties in PID control and tuning. Sequential learning methods (e.g., BPPO) are competitive against some diffusion-based approaches but sacrifice Target Loss performance and underperform compared to global trajectory planning methods in other systems. Diffusion-based methods demonstrate superior overall performance, as they better capture long-term dependencies in system dynamics compared to traditional and iterative sequential methods, avoiding myopic failure modes and facilitating global optimization of long-term dynamics.

Due to space constraints, Appendices~\ref{sup:numeric} and \ref{sup:visual} provide detailed numerical results, including standard errors, Pareto frontier comparisons, and control dynamics visualizations. A comparison with a data-driven MPC in Appendix~\ref{sup:mpc} further underscores the advantages of our global planning approach. While our main experiments assume full-state observability, we validate SEDC's robustness to significant observation noise in Appendices~\ref{app:noise}. Furthermore, Appendices~\ref{app:jellyfish} validate the method's effectiveness on higher-dimensional physics systems.

\begin{table}[t]
\vspace{-2pt}
\centering
\small
\caption{Ablation performance across datasets. \textbf{Target Loss} at 10\% and 100\% training data is reported. Best/second-best/worst in each row are in \textbf{bold}/\underline{underlined}/\textit{italics}.}
\label{tab:ablation_main}
\small
\begin{tabular}{@{}l@{\;}l@{\;}|@{\;}c@{\quad}c@{\quad}c@{\quad}c@{}}
\toprule
System & Ratio & Ours & w/o DSD & w/o DMD & w/o GSF \\
\midrule
Burgers & 10\% & \textbf{1.74e-4} & \textit{1.00e-3} & \underline{3.78e-4} & 6.67e-4 \\
        & 100\% & \textbf{9.80e-5} & \textit{8.71e-4} & \underline{2.28e-4} & 2.62e-4 \\
\cmidrule{1-6}
Kuramoto & 10\% & \textbf{1.12e-5} & \textit{4.15e-3} & \underline{5.21e-5} & 4.77e-5 \\
         & 100\% & \textbf{8.90e-6} & \textit{5.43e-3} & \underline{1.76e-5} & 3.88e-5 \\
\cmidrule{1-6}
IP       & 10\% & \textbf{6.21e-4} & \textit{1.58e-3} & \underline{1.10e-3} & 2.00e-3 \\
         & 100\% & \textbf{3.49e-4} & \textit{1.37e-3} & \underline{6.64e-4} & 7.85e-4 \\
\bottomrule
\end{tabular}
\vspace{-1em}
\end{table}

\subsection{Sample Efficiency}

To evaluate the sample efficiency of diffusion-based methods, we conducted experiments on all the systems using varying proportions of the full training dataset. Specifically, we trained models using 1\%, 5\%, 10\%, 20\%, and 100\% of the available data and assessed their performance using the Target Loss metric on a held-out test set. 
Figure~\ref{fig:efficiency} demonstrates our method's superior performance in controlling Burgers and Kuramoto systems compared to state-of-the-art baselines. In all systems, our approach achieves significantly lower target loss values across all training data percentages. Most notably, with only 10\% of the training data, our method attains a target loss of 1.71e-4 for Burgers, 1.12e-5 for Kuramoto, and 6.35e-4 for Inverse Pendulum, matching(-5.5\% in Burgers) or exceeding(+36.4\% in Kuramoto and +1.2\% in Inverse Pendulum) the performance of best baseline methods trained on the complete dataset. This indicates our method can achieve state-of-the-art performance while requiring only 10\% of the training samples.

Among the baselines, DiffPhyCon's complex training requirement (Sec.~\ref{rw}) makes it particularly sample-inefficient. While AdaptDiffuser's self-tuning improves upon DecisionDiffuser, SEDC's advantage stems from its efficient dynamics learning and a filter-free finetuning strategy. Unlike AdaptDiffuser, our GSF mechanism integrates all guided trajectories without discriminator filtering, promoting data diversity and a better exploration-exploitation balance.

\begin{figure}[t]
    \centering
    \begin{subfigure}[t]{0.39\linewidth}
        \centering
        \includegraphics[width=\linewidth]{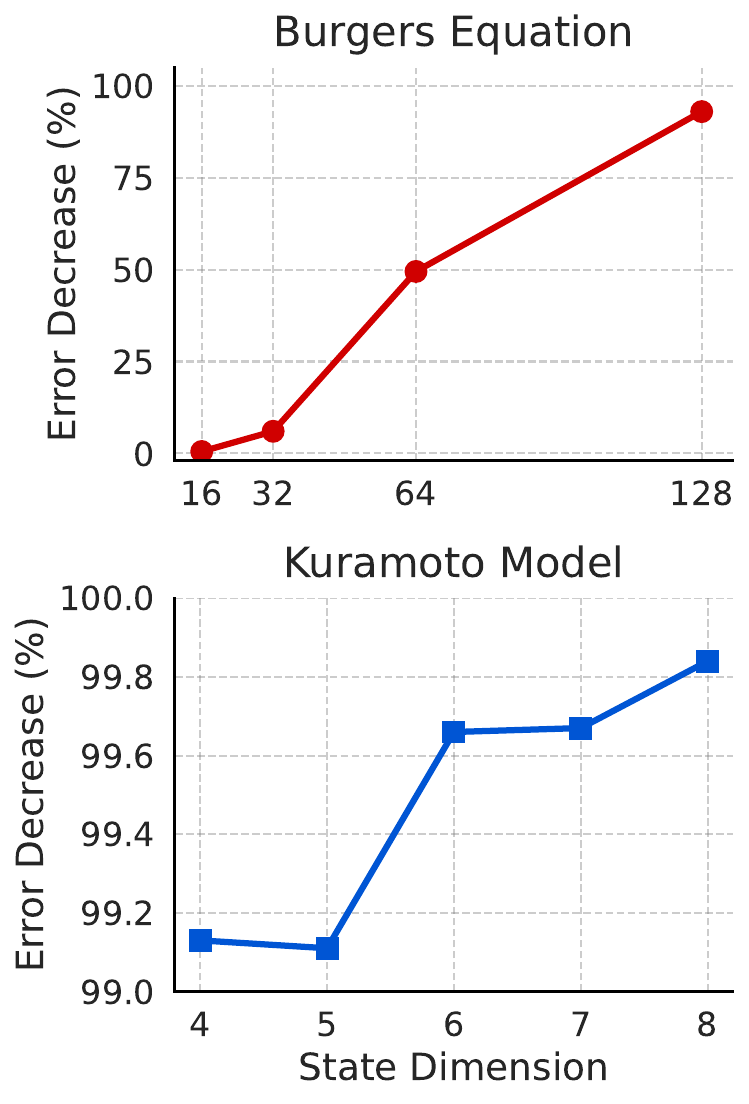}
        \caption{\phantom{.}}
        \vspace{-0.5em}
        \label{fig:dsd_comparison}
    \end{subfigure}
    \hfill
    \begin{subfigure}[t]{0.60\linewidth}
        \centering
        \includegraphics[width=\linewidth]{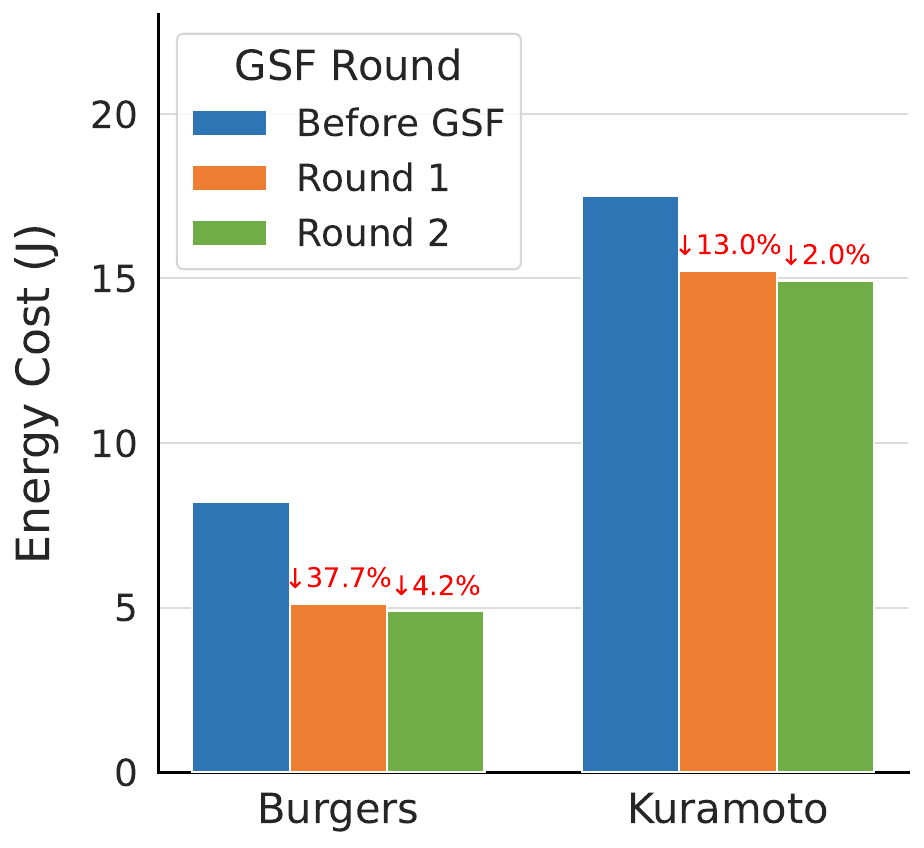}
        \caption{\phantom{.}}
        \vspace{-0.5em}
        \label{fig:gsf_ablation}
    \end{subfigure}
    \caption{Ablation analysis of SEDC components. \textbf{(a)} Target Loss decrease rate of Decoupled State Diffusion (DSD), measured as the reduction relative to \textit{w/o DSD} across state dimensions in Burgers and Kuramoto. \textbf{(b)} Energy cost ($J$) reduction over Guided Self-finetuning (GSF) rounds, with most gains in the first round and subsequent rounds indicating convergence.}
    \label{fig:ablation_dsd_gsf}
\vspace{-1.5em}
\end{figure}

\subsection{Ablation Study}

\textbf{Overall ablation study.} We explore the main performance against each ablation of the original SEDC. Specifically, \textit{w/o DSD} removes the inverse dynamics, unifying the diffusion of system state and control input, i.e. $\mathbf{x}=[\mathbf{u},\mathbf{y}]$. Therefore, the diffusion model is required to simultaneously capture the temporal information and implicit dynamics of the control and system trajectory. Note that the inpainting mechanism and gradient guidance are retained. \textit{w/o DMD} removes the decomposition design, resulting in a single 1-D Unet structure as the denoising network, following DecisionDiff. Finally, \textit{w/o GSF} reports the performance without guided self-finetuning process, which means the model only uses the original dataset to train itself. To show the sample-efficiency performance, we also investigate the results under less amount of training sample(10\%). For \textit{w/o DMD} and \textit{w/o DSD}, we adjust the number of trainable parameters at a comparable level against the original version.

Table~\ref{tab:ablation_main} shows the Target Loss performance of different ablations of SEDC
across multiple datasets and different training sample ratios. As can be seen, removing any component leads to a certain decrease in performance, whether the training data is limited or not, demonstrating the effectiveness of each design. The most significant performance drops are often observed in \textit{w/o DSD}, highlighting the importance of explicit learning of dynamics in complex systems. \textit{w/o DMD} exhibits the lowest decline across the three systems. This is because the single-Unet-structured denoising network can already capture the nonlinearity to some extent, but not as good as the proposed decomposition approach. with 10\% of training data, removing individual components still led to noticeable performance degradation, and the patterns consistent with the full dataset results. This demonstrates that our designs remain effective in low-data scenarios.

\textbf{Effectiveness of DSD.} 
To evaluate DSD's effectiveness against the curse of dimensionality, we compared the performance of original and \textit{w/o DSD} models across Kuramoto systems with dimensions ranging from $N=4$ to $N=8$. Experimental results (Figure~\ref{fig:dsd_comparison}) 
show that performance degradation from \textit{w/o DSD} increases proportionally with system dimensionality, demonstrating DSD's enhanced effectiveness in higher-dimensional systems and validating its capability to address dimensionality challenges. Further, we can observe that the decrease changes more rapidly with increasing dimensionality compared to Kuramoto, indicating that DSD exhibits heightened sensitivity to dimensionality in higher-dimensional systems. We also investigate the effectiveness of dynamical learning in Appendix \ref{app:consistency}.

\begin{table}[t]
\caption{Performance degradation using different denoiser output with varying nonlinearity strength $\gamma$ in the Kuramoto system. }
\centering
\small
\begin{tabular}{cccc}
\toprule
$\gamma$ & 1 & 2 & 4 \\
\midrule
$\mathbf{O}_1 + \mathbf{O}_2$ & 8.90e-6 & 2.78e-5 & 3.89e-5 \\
$\mathbf{O}_1$ & 1.42e-5 & 4.73e-5 & 8.52e-5 \\
Dec. (\%) & 37.3 & 41.2 & 54.3 \\
\bottomrule
\end{tabular}
\label{tab:dmd}
\vspace{-1.75em}
\end{table}

\textbf{Effectiveness of DMD.} To investigate the contribution of DMD's dual-Unet architecture to nonlinearity learning, we conducted experiments on the Kuramoto system with varying degrees of nonlinearity (controlled by the coefficient $\gamma\in\{1,2,4\}$ of the nonlinear sinusoidal term, where larger values indicate stronger nonlinearity). 
We compared the performance between using only the linear intermediate output ($\hat{\mathbf{x}}_0=\mathbf{O}_1$) of the denoising network and the original nonlinear output ($\hat{\mathbf{x}}_0=\mathbf{O}_1+\mathbf{O}_2$) in terms of Target Loss. The Dec. indicates the reduction in target loss achieved by nonlinear output $\mathbf{O}_1 + \mathbf{O}_2$ compared to linear output $\mathbf{O}_1$. As shown in Table~\ref{tab:dmd}, the magnitude of loss reduction increases proportionally with the nonlinearity strength $\gamma$, indicating that the quadratic term exhibits enhanced capability in capturing nonlinear dynamics as the system's nonlinearity intensifies. This comparison confirms that the nonlinear branch $\mathbf{O}_2$ is essential for capturing dynamics as nonlinearity ($\gamma$) increases. The superiority of the decomposition architecture itself is further evidenced in Table~\ref{tab:ablation_main}, where \textit{w/o DMD} shows consistently higher target loss. We also show that DMD is sufficient for modeling even higher-order dynamics in Appendix~\ref{app:dmd_sufficiency}.

\textbf{Effectiveness of GSF.} 
To validate GSF's ability to refine the control law towards lower energy cost, we tracked test-set energy consumption across finetuning rounds. As shown in Figure~\ref{fig:gsf_ablation}, the first GSF round yields a dramatic improvement over the initial model, reducing energy by 37.7\% (Burgers) and 13.0\% (Kuramoto). Subsequent rounds offer minor refinements (4.2\% and 2.0\% respectively), indicating convergence to a near-optimal law. This confirms GSF effectively guides the model beyond its initial suboptimal training data to discover more energy-efficient control solutions.


\begin{figure}[t]
    \centering
    \begin{minipage}{0.48\linewidth}
        \centering
        \includegraphics[width=\linewidth]{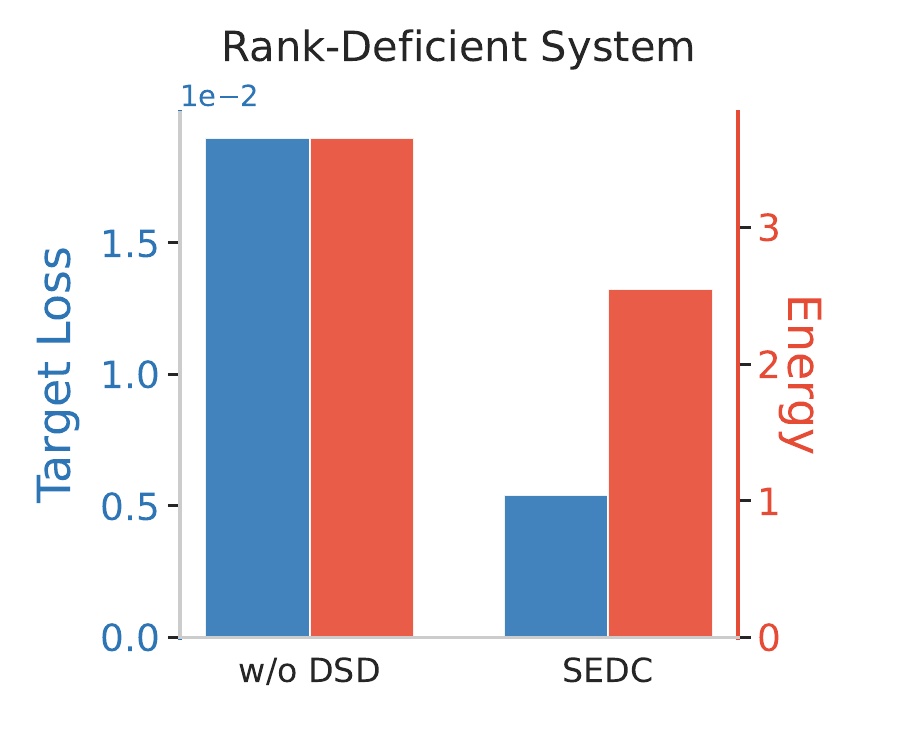}
        \caption{\small Comparison on a rank-deficient system showing SEDC's robustness to multi-modal control.}
        \label{fig:case_rank}
    \end{minipage}
    \hfill 
    \begin{minipage}{0.48\linewidth}
        \centering
        \includegraphics[width=\linewidth]{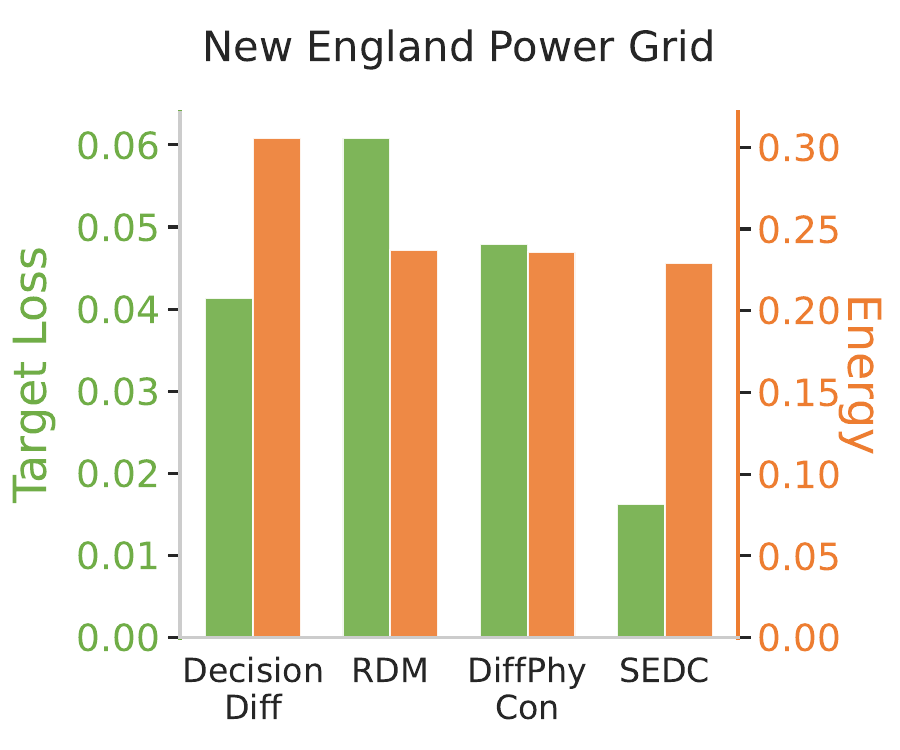}
        \caption{\small Stabilization performance on the 39-bus New England Grid comparing SEDC with diffusion baselines.}
        \label{fig:case_grid}
    \end{minipage}
    \vspace{-1.0em}
\end{figure}

\textbf{Robustness to Non-invertibility.}
\label{sec:non_invertible}
A critical question is whether DSD fails when multiple controls map to the same state transition (multi-modality). To investigate this, we tested SEDC against the joint-diffusion baseline (\textit{w/o DSD}) on a rank-deficient system where control is non-unique. Details can be found in Appendix~\ref{app:non_invertible}.
As shown in Figure~\ref{fig:case_rank}, SEDC outperforms the baseline by reducing Target Loss by over 70\%. We attribute this success to task decomposition: joint baselines struggle to model the complex multi-modal joint distribution $p(\mathbf{y}, \mathbf{u})$, often leading to mode averaging. 
In contrast, SEDC first generates a coherent state trajectory $\mathbf{y}_{0:T}$. Conditioned on this valid path, the inverse model $f_\phi$ transforms the intractable joint generation problem into a simplified supervised task: identifying \textit{a} feasible control (e.g., the smoothest one) that satisfies the transition.

\subsection{Case Study}
\label{sec:case_study}
\textbf{Real-world Power Grid.}
To validate scalability on real-world infrastructure, we applied SEDC to the 39-bus New England power grid (Swing Equation)~\citep{baggio2021data}, a real-world benchmark for post-fault stabilization (setup details in Appendix~\ref{app:swing}).
As shown in Figure~\ref{fig:case_grid}, SEDC achieves a 60.5\% reduction in Target Loss compared to the strongest baseline (DecisionDiffuser) while maintaining lower energy cost.
This significant margin demonstrates SEDC's superior ability to capture the highly coupled nonlinearities of power networks, where our global planning approach avoids the compounding errors typical of myopic methods in recovering synchronous states.

\section{Conclusion}
In this paper, we presented SEDC, a novel sample-efficient diffusion-based framework for complex nonlinear system control. By synergistically integrating Decoupled State Diffusion (DSD), Dual-Mode Decomposition (DMD), and Guided Self-finetuning (GSF), SEDC achieves superior control performance with remarkable data efficiency. Our experiments show that SEDC can match state-of-the-art accuracy using just 10\% of the training data. The framework's robustness is further validated through rigorous testing on high-dimensional PDEs and real-world systems, confirming the broad applicability and effectiveness of our design principles. These results mark a significant advancement in developing practical and sample-efficient solutions for complex system control.








\section*{Impact Statement}
This paper presents work whose goal is to advance the field of Machine
Learning. There are many potential societal consequences of our work, none
which we feel must be specifically highlighted here.


\bibliography{example_paper}
\bibliographystyle{icml2026}

\newpage
\appendix
\onecolumn
\section{Algorithm form of SEDC}
\label{sup:alg}
\begin{algorithm}[H]
\caption{SEDC: Training and finetuning}
\KwIn{Initial dataset $\mathcal{D}_0$, diffusion steps $K$, guidance strength $\lambda$, 
    self-finetuning rounds $R$, forward dynamics $f_{\text{forward}}$}
\KwOut{Optimized trajectory $\mathbf{y}_{0:T}^0$, controls $\mathbf{u}_{0:T}^0$}

\SetKwProg{Fn}{Function}{}{}
\Fn{Initial Training($\mathcal{D}_0$)}{
    \While{not converged}{
        Sample batch $(\mathbf{y}_{0:T}, \mathbf{u}_{0:T}) \sim \mathcal{D}_0$ \\
        Sample $k \sim \mathcal{U}\{1,...,K\}$, $\epsilon \sim \mathcal{N}(0,I)$ \\
        Corrupt states: $\mathbf{y}^k = \sqrt{\bar{\alpha}^k}\mathbf{y} + \sqrt{1-\bar{\alpha}^k}\epsilon$ \\
        Predict clean states: $\hat{\mathbf{y}}^k = \mathcal{G}_\theta(\mathbf{y}^k, k, \mathbf{y}_0^*, \mathbf{y}_f)$ \\
        Predict controls: $\tilde{\mathbf{u}}_t = f_\phi({\mathbf{y}}_t, {\mathbf{y}}_{t+1})$ \\
        Compute losses: 
        \quad $L_{\text{diff}} = \|\mathbf{y} - \hat{\mathbf{y}}^k\|^2$ \\
        \quad $L_{\text{inv}} = \|\mathbf{u}_t - \hat{\mathbf{u}}_t\|^2$ \\
        Update $\theta, \phi$ with $\nabla(L_{\text{diff}} + L_{\text{inv}})$
    }
}

\For{$r = 1$ \textbf{to} $R$}{
    \textbf{Guided Data Generation:} \\
    Initialize $\mathbf{y}^K \sim \mathcal{N}(0, I)$, sample $(\mathbf{y}_0^*,\mathbf{y}_f)\sim \mathcal{D}_{r-1}$ \\
    \For{$k = K$ \textbf{downto} $1$}{
        Predict $\hat{\mathbf{y}}^k = \mathcal{G}_\theta(\mathbf{y}^k, k, \mathbf{y}_0^*, \mathbf{y}_f)$ \\
        Compute gradient: $g = \nabla_{\mathbf{y}^k} J(\tilde{\mathbf{y}}^k,\hat{\mathbf{u}}^k)$, where $\tilde{\mathbf{u}}_t^k = f_\phi(\hat{\mathbf{y}}_t^k, \hat{\mathbf{y}}_{t+1}^k)$ \\
        Adjust mean: $\mu_\theta = \mu_\theta^{\text{(base)}} - \lambda \Sigma^k g$ \\
        Sample $\mathbf{y}^{k-1} \sim \mathcal{N}(\mu_\theta, \Sigma^k I)$ \\
        Enforce constraints: $\mathbf{y}^{k-1}_0 \leftarrow \mathbf{y}_0^*$, $\mathbf{y}^{k-1}_T \leftarrow \mathbf{y}_f$
    }
    Recover controls: $\tilde{\mathbf{u}}_t^0 = f_\phi(\mathbf{y}_t^0, \mathbf{y}_{t+1}^0)$ \\
    \textbf{System Interaction:} \\
    Generate $\tilde{\mathbf{y}}_{[0:T]}^0 = f_{\text{forward}}(\tilde{\mathbf{u}}_{[0:T]}^0, \mathbf{y}_0^*)$ \\
    Augment dataset: $\mathcal{D}_r = \mathcal{D}_{r-1} \cup \{(\tilde{\mathbf{y}}_{[0:T]}^0, \tilde{\mathbf{u}}_{[0:T]}^0\}$ \\
    
    \textbf{Adaptive Fine-tuning:} \\
    \While{validation loss decreases}{
        Sample batch from $\mathcal{D}_r$ \\
        Perform training steps as in Initial Training
    }
}

\Return Optimized $\theta,\phi$

\textit{(Test process follows guided data generation with test conditions $(\mathbf{y}_0^*,\mathbf{y}_f)$ provided.)}

\end{algorithm}

\section{Detailed System and Dataset Description}
\label{app:b}
\subsection{Overall Introduction}
Our benchmark selection (Inverted Pendulum, Kuramoto, and Burgers) follows established control systems research practice, chosen for real-world relevance and diverse nonlinearity and complexity:
\begin{itemize}
    \item Inverted Pendulum: state 2, control 1, timestep 128
    \item Kuramoto: state 8, control 8, timestep 15
    \item Burgers: state 128, control 128, timestep 10
\end{itemize}

\subsection{Burgers Dynamics}

The Burgers' equation is a governing law occurring in various physical systems. We consider the 1D Burgers' equation with the Dirichlet boundary condition and external control input $\mathbf{u}(t,x)$:

$$
\begin{cases}
\frac{\partial y}{\partial t} = -y \cdot \frac{\partial y}{\partial x} + \nu \frac{\partial^2 y}{\partial x^2} + \mathbf{u}(t,x) & \text{in } [0,T] \times \Omega \\
y(t,x) = 0 & \text{on } [0,T] \times \partial\Omega \\
y(0,x) = y_0(x) & \text{in } \{t=0\} \times \Omega
\end{cases}
$$

Here $\nu$ is the viscosity parameter, and $y_0(\mathbf{x})$ is the initial condition. Subject to these equations, given a target state $y_d(x)$, the objective of control is to minimize the control error $\mathcal{J}_\text{actual}$ between $y_T$ and $y_d$, while constraining the energy cost $\mathcal{J}_\text{energy}$ of the control sequence $\mathbf{u}(t,x)$.

We follow instructions in~\citet{wei2024generative} to generate a 1D Burgers’ equation dataset. Specifically, for numerical simulation, we discretized the spatial domain [0,1] and temporal domain [0,1] using the finite difference method (FDM). The spatial grid consisted of 128 points, while the temporal domain was divided into 10000 timesteps. We initiated the system with randomly sampled initial conditions and control inputs drawn from specified probability distributions. This setup allowed us to generate 90000 trajectories for training and 50 trajectories for testing purposes.

\subsection{Kuramoto Dynamics}
The Kuramoto model is a paradigmatic system for studying synchronization phenomena. We considered a ring network of $N=8$ Kuramoto oscillators. The dynamics of the phases (states) of oscillators are expressed by:

\begin{equation}
    \dot\theta_{i,t} = \omega + \gamma(\sin(\theta_{i-1,t-1}-\theta_{i,t-1}) + \sin(\theta_{i+1,t-1}-\theta_{i,t-1}))+u_{i,t-1}, \quad i=1,2,...,N.
\end{equation}

For the Kuramoto model, we generated 20,000 samples for training and 50 samples for testing. The initial phases were sampled from a Gaussian distribution $\mathcal{N}(0,I)$, and the random intervention control signals were sampled from $\mathcal{N}(0,2I)$. The system was simulated for $T=16$ time steps with $\omega=0$, following~\citet{baggio2021data}. The resulting phase observations and control signals were used as the training and test datasets.

\subsection{Inverted Pendulum Dynamics}
The inverted pendulum is a classic nonlinear control system. The dynamics can be represented by:

$$
\frac{d^2\theta}{dt^2} = \frac{g}{L}\sin(\theta) - \frac{\mu}{L}\frac{d\theta}{dt} + \frac{1}{mL^2}u
$$

where $\theta$ is the angle from the upward position, and $u$ is the control input torque. The system parameters are set as: gravity $g=9.81$ m/s², pendulum length $L=1.0$ m, mass $m=1.0$ kg, and friction coefficient $\mu=0.1$.

To generate the training dataset, we simulate 90,000 trajectories for training and 50 for testing with 128 time steps each, using a time step of 0.01s. For each trajectory, we randomly sample initial states near the unstable equilibrium point with $\theta_0 \sim \mathcal{U}(-1,1)$ and $\dot{\theta}_0 \sim \mathcal{U}(-1,1)$, and generate control inputs from $u \sim \mathcal{U}(-0.5,0.5)$. The resulting dataset contains the state trajectories and their corresponding control sequences.
\section{Implementation Details}
\label{sup:imp}
\subsection{Implementation of SEDC}
In this section, we describe various architectural and hyperparameter details:
\begin{itemize}
    \item The temporal U-Net~(1D-Unet) \citep{janner2022planning} in the denoising network consists of a U-Net structure with $4$ repeated residual blocks. Each block comprises two temporal convolutions, followed by group normalization, and a final Mish nonlinearity. The channel dimensions of the downsample layers are ${1,2,4}*state dimension$. Timestep embedding is produced by a Sinusoidal Positional Encoder, following a 2-layer MLP, and the dimension of this embedding is 32. The dimension of condition embedding is the same as the system state dimension.
    \item We represent the inverse dynamics $f_\phi$ with an autoregressive model with 64 hidden units and ReLU activations. The model autoregressively generates control outputs along the control dimensions.
    \item We train $\mathbf{x}_\theta$ and $f_\phi$ using the Adam optimizer with learning rates from \{1e-3, 5e-3, 1e-4\}. The exact choice varies by task. Moreover, we also use a learning rate scheduler with step factor=0.1. Training batch size is 32.
    \item We use $K=128$ diffusion steps.
    \item We use a guidance scale $\lambda\in\{0.01,0.001,0.1\}$ but the exact choice varies by task.
\end{itemize}

\subsection{Training and Inference Time Analysis}

\begin{table}[h]
    \centering
    \caption{Approximate Training Time Comparison of Different Models on Various Datasets (in hours)}
    \label{tab:training_time}
    \begin{tabular}{@{}lccccc@{}}
    \toprule
    Dataset/System & DecisionDiffuser & RDM & DiffPhyCon & AdaptDiffuser & SEDC \\
    \midrule
    Burgers & 2.5 & 2.5 & 3.0 & 2.5 & 2.5 \\
    Kuramoto & 1.5 & 1.5 & 1.5 & 1.0 & 1.0 \\
    IP & 1.0 & 1.0 & 1.5 & 1.0 & 0.5 \\
    \bottomrule
    \end{tabular}
\end{table}

\begin{table}[h]
    \centering
    \caption{Approximate Inference Time Comparison of Different Models on Various Datasets (in seconds)}
    \label{tab:inference_time}
    \begin{tabular}{@{}lccccc@{}}
    \toprule
    Dataset/System & DecisionDiffuser & RDM & DiffPhyCon & AdaptDiffuser & SEDC \\
    \midrule
    Burgers & 3.0 & 4.0 & 6.0 & 4.0 & 4.0 \\
    Kuramoto & 1.0 & 1.5 & 2.0 & 1.5 & 1.5 \\
    IP & 0.5 & 1.0 & 1.0 & 0.5 & 0.5 \\
    \bottomrule
    \end{tabular}
\end{table}

The diffusion-based methods are trained on single NVIDIA GeForce RTX 4090 GPU. We evaluate the training and inference time of all the diffusion-based methods evaluated in the experiment session.
As shown in Table \ref{tab:training_time}, we compare the training efficiency of different models across various datasets. DiffPhyCon consistently shows longer training times compared to other methods, because it requires training two models that learn the joint distribution and the prior distribution respectively, increasing its training time consumption. The training times of DecisionDiffuser, RDM, and AdaptDiffuser are generally comparable, while SEDC demonstrates relatively efficient training performance across most datasets. This may be because of the proposed designs that not only improve sample efficiency but also improve learning efficiency.

The inference time comparison in Table \ref{tab:inference_time} reveals that DiffPhyCon requires longer execution time compared to other models, because it needs to sample from two learned distributions in the denoising process. RDM achieves relatively slower inference speeds than DecisionDiffuser, AdaptDiffuser, and SEDC, because RDM replans during inference, increasing planning time. Notably, all models exhibit shorter training and inference times on the IP dataset, suggesting the influence of system complexity on computational efficiency. 

\section{Baselines Description}
\label{sup:baselines}
\subsection{PID}

PID (Proportional-Integral-Derivative) control is a classical feedback control methodology that has been widely adopted in industrial applications. The control signal is generated by computing the weighted sum of proportional, integral, and derivative terms of the error. The control law can be expressed as:

$$
u(t) = K_p e(t) + K_i \int_0^t e(\tau)d\tau + K_d \frac{d}{dt}e(t)
$$

While PID controllers exhibit robust performance and require minimal system modeling, their effectiveness may be compromised when dealing with highly nonlinear or time-varying systems, necessitating frequent parameter tuning.

\subsection{BC, BPPO}

Behavior Cloning (BC) represents a supervised imitation learning paradigm that aims to learn a direct mapping from states to actions by minimizing the deviation between predicted actions and expert demonstrations. Despite its implementation simplicity and sample efficiency, BC suffers from distributional shift, where performance degradation occurs when encountering states outside the training distribution. The objective function can be formulated as:

$$
L_{BC}(\theta) = \mathbb{E}_{(s,a)\sim \mathcal{D}} [-\log \pi_\theta(a|s)]
$$

where $\mathcal{D}$ denotes the expert demonstration dataset.

Behavior-guided PPO (BPPO) presents a hybrid approach that integrates behavior cloning with Proximal Policy Optimization. By incorporating a behavioral cloning loss term into the PPO objective, BPPO facilitates more efficient policy learning while maintaining the exploration capabilities inherent to PPO. The composite objective function is defined as:

$$
L_{BPPO}(\theta) = L_{PPO}(\theta) + \alpha L_{BC}(\theta)
$$

where $\alpha$ serves as a balancing coefficient between the PPO and BC objectives.

Each method exhibits distinct characteristics: BC demonstrates effectiveness when abundant high-quality expert demonstrations are available. BPPO leverages the synergy between expert knowledge and reinforcement learning for complex control scenarios. 

\subsection{Diffusion-based methods}

\begin{itemize}
    \item \textbf{DecisionDiffuser:} \\
        A novel approach that reformulates sequential decision-making as a conditional generative modeling problem rather than a reinforcement learning task. The core methodology involves modeling policies as return-conditional diffusion models, enabling direct learning from data without dynamic programming. The model can be conditioned on various factors including constraints and skills during training.

    \item \textbf{DiffPhyCon:} \\
        A diffusion-based method for controlling physical systems that operates by jointly optimizing a learned generative energy function and predefined control objectives across entire trajectories. The approach incorporates a prior reweighting mechanism to enable exploration beyond the training distribution, allowing the discovery of diverse control sequences while respecting system dynamics.

    \item \textbf{AdaptDiffuser:} \\
        An evolutionary planning framework that enhances diffusion models through self-evolution. The method generates synthetic expert data using reward gradient guidance for goal-conditioned tasks, and employs a discriminator-based selection mechanism to identify high-quality data for model fine-tuning. This approach enables adaptation to both seen and unseen tasks through continuous model improvement.

    \item \textbf{RDM:} \\
        A replanning framework for diffusion-based planning systems that determines replanning timing based on the diffusion model's likelihood estimates of existing plans. The method introduces a mechanism to replan existing trajectories while maintaining consistency with original goal states, enabling efficient bootstrapping from previously generated plans while adapting to dynamic environments.
\end{itemize}

\section{Detailed Results of Figure 2}
\label{sup:numeric}

\begin{table*}[h]
\small
\centering
\caption{\textbf{Performance comparison of different models across three datasets.} TL (Target Loss) and $J$ (Energy) are reported, with lower values indicating better performance for both metrics. We report the mean and the standard error over 5 random seeds. Following previous work(e.g. \citet{ajay2022conditional}), we highlight the best-performed results in \textbf{bold}.}
\label{tab:main_comparison}
\begin{tabular}{l|cc|cc|cc}
\toprule
\multirow{2}{*}{\textbf{Model}} & \multicolumn{2}{c|}{\textbf{Burgers}} & \multicolumn{2}{c|}{\textbf{Kuramoto}} & \multicolumn{2}{c}{\textbf{IP}} \\
\cmidrule{2-7}
& TL & $J$ & TL & $J$ & TL & $J$ \\
\midrule
PID & 1.30e-1 & 6.56 & 7.99e-1 & 30.35 & 8.64e-3 & 2.28e-1 \\
BPPO & 5.90e-4 & 9.72 & 1.56e-4 & 26.64 & 3.63e-3 & 4.16e-3 \\
BC & 4.78e-4 & 10.73 & 1.52e-4 & 27.59 & 3.63e-3 & 4.20e-3 \\
DecisionDiff & 2.46e-4 & 5.18 & 3.88e-5 & 27.48 & \textbf{6.65e-4} & \textbf{9.00e-4} \\
RDM & 2.70e-4 & 7.01 & 4.60e-4 & 29.03 & 7.85e-4 & 3.38e-3 \\
DiffPhyCon & \textbf{1.62e-4} & \textbf{5.15} & 4.80e-4 & 18.72 & \textbf{6.63e-4} & \textbf{1.99e-3} \\
AdaptDiffuser & 2.28e-4 & \textbf{4.65} & \textbf{1.76e-5} & 26.23 & 8.64e-4 & 5.49e-3 \\
Ours & \textbf{9.80{\scriptsize±5.6}e-5} & \textbf{5.01{\scriptsize±0.6}} & \textbf{8.90{\scriptsize±3.1}e-6} & \textbf{14.90{\scriptsize±0.8}} & \textbf{3.49{\scriptsize±2.6}e-4} & \textbf{8.90{\scriptsize±0.9}e-4} \\
\bottomrule
\end{tabular}
\end{table*}

\begin{figure}[h]
    \centering
    \includegraphics[width=\linewidth]{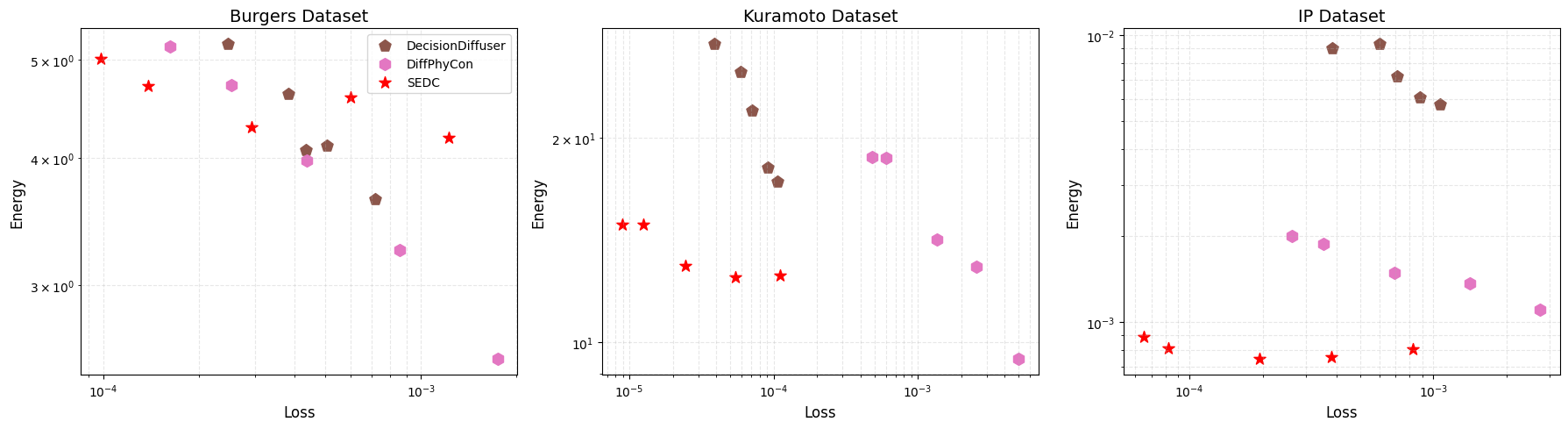}
    \caption{Pareto frontier of target loss and energy cost $J$ across different datasets of our method and two SOTA baselines DecisionDiffuser and DiffPhyCon. The closer the data point is to the bottom left, the better the performance.}
    \label{fig:pareto}
\end{figure}

We leverage 2-D plots in the main paper to better illustrate the performance comparison of all the methods. Here we provide the provides the corresponding numerical results in detail in Table~\ref{tab:main_comparison} and pareto frontiers in Fig~\ref{fig:pareto}. Results confirm our method maintains competitive performance across all datasets. 

\begin{figure}[h]
    \centering
    \includegraphics[width=0.8\linewidth]{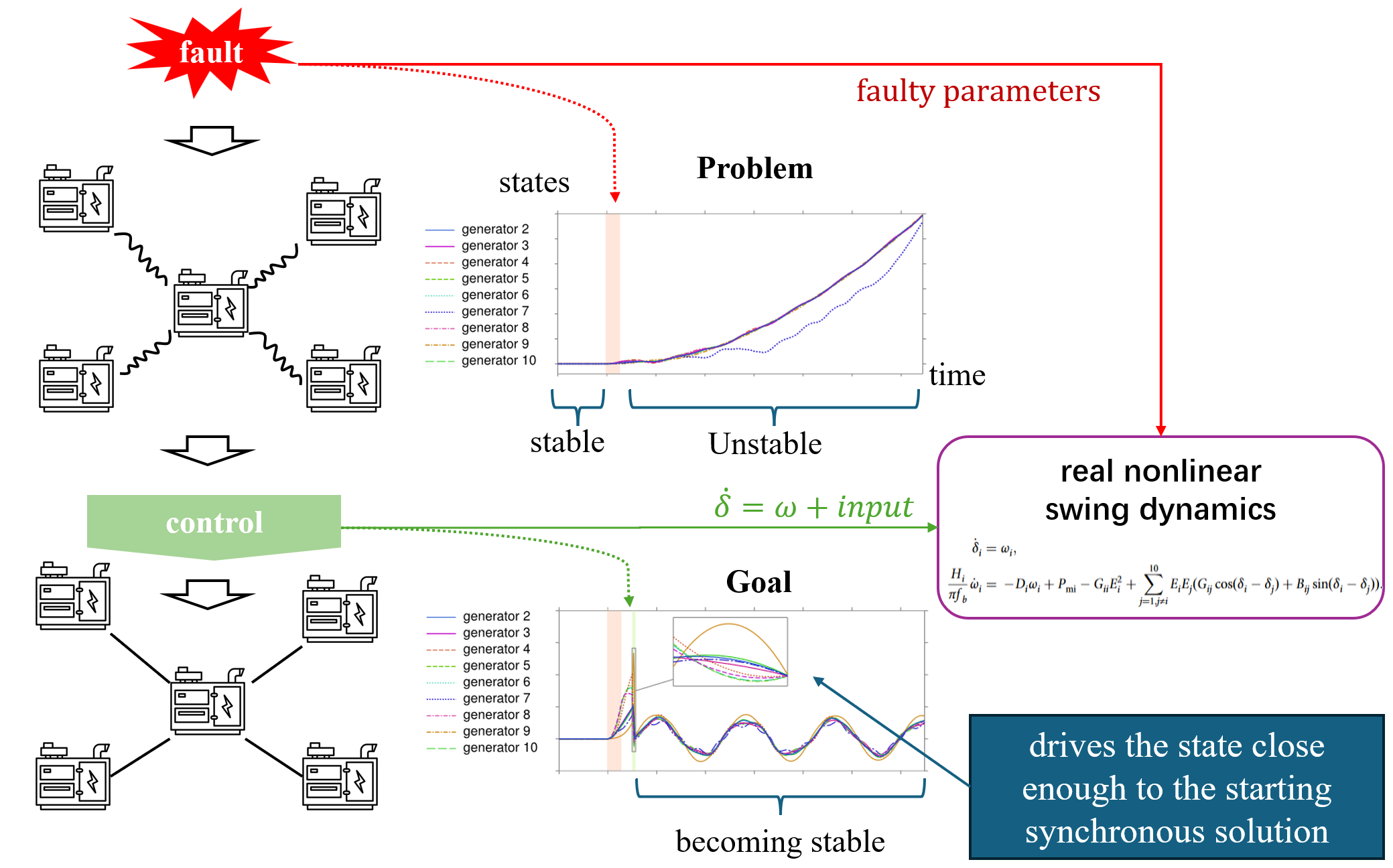}
    \caption{Illustration of the power grid swing dynamics control task. The system simulates a recovery scenario following a line fault, where the objective is to compute an optimal control sequence that drives the generators from a disturbed, un-synchronized state back to steady-state operation (zero deviation). Several component of this figure are adapted from~\citet{baggio2021data}.}
    \label{fig:swing_diagram}
\end{figure}

\section{Power Grid Swing Dynamics Details}
\label{app:swing}

In Section~\ref{sec:case_study}, we presented comparative results on the power grid swing dynamics. Here we provide the specific experimental configuration.

\textbf{System Description.}
The Swing Equation system models the rotational dynamics of synchronous generators in a power network. We utilize the New England 10-generator 39-bus system. The state space consists of the rotor angles and frequencies of the generators, resulting in a high-dimensional nonlinear system.

\textbf{Task Setup.}
Following the experimental design in~\citet{baggio2021data}, the control task simulates a recovery scenario following a line fault. The process is illustrated in Figure~\ref{fig:swing_diagram}. 
The objective is to compute an optimal control sequence that drives the generators from a disturbed, un-synchronized state back to the steady-state operation (zero deviation).
The trajectories are sampled with \textbf{18 state variables} and \textbf{9 control inputs} over a horizon of \textbf{32 timesteps}. This setup incorporates realistic physical parameters and disturbances, serving as a rigorous test for controlling coupled nonlinear dynamics.

\section{Supplementary Experiments}
\label{sup:sup}

\subsection{Comparison against MPC}
\label{sup:mpc}
\begin{table*}[h]
\small
\setlength{\tabcolsep}{3pt}
\centering
\caption{Comparison of SEDC (Ours) and Model Predictive Control (MPC) across three control tasks. Results show target loss (MSE), control cost (J), and inference time in seconds. We implemented the MPC that uses a neural network architecture with residual connections to learn system dynamics from data, then solves finite-horizon optimization problems at each timestep using the gradient of the summation of target loss and control energy, which iteratively refines the sampling distribution toward optimal control sequences. The control horizons of Kuramoto,Burgers and IP are 15, 10 and 128.}
\begin{tabular}{@{}l@{\hspace{0.5em}}ccc@{\hspace{1em}}ccc@{\hspace{1em}}ccc@{}}
\toprule
\multirow{2}{*}{Model} & \multicolumn{3}{c}{Kuramoto} & \multicolumn{3}{c}{Burgers} & \multicolumn{3}{c}{IP} \\
\cmidrule(lr){2-4} \cmidrule(lr){5-7} \cmidrule(l){8-10}
 & Target loss & J & Time(s) & Target loss & J & Time(s) & Target loss & J & Time(s) \\
\midrule
MPC & 9.10e-03 & 0.34 & $\sim$50 & 1.89e-01 & 0.001 & $\sim$30 & 4.33e-04 & 1.01e-03 & $>$1000 \\
Ours & 8.90e-06 & 14.9 & $\sim$1.5 & 9.80e-05 & 5.01 & $\sim$2.5 & 3.49e-04 & 8.90e-04 & $\sim$0.5 \\
\bottomrule
\end{tabular}
\label{tab:mpc}
\end{table*}

We additionally compare SEDC with data-driven Model Predictive Control (MPC). We implement MPC in a data-driven way, where we train an MLP for the forward model.  Results in Table~\ref{tab:mpc} show MPC achieves higher target losses across all tasks, likely due to error accumulation. MPC also has significantly longer inference times (e.g. >1000 vs. 0.5) that increase with control horizon. SEDC directly maps initial/target states to complete control trajectories, avoiding compounding errors and reducing computation time.

\subsection{Validation on High-Dimensional 2D PDE Control: Jellyfish Locomotion}
\label{app:jellyfish}

To address the critical question of our method's scalability and effectiveness on truly high-dimensional control problems, we conducted additional experiments on a challenging 2D PDE control benchmark: the locomotion of a jellyfish. This task represents a state-of-the-art challenge in data-driven control of physical systems, involving complex fluid-solid interactions governed by the 2D incompressible Navier-Stokes equations~\citep{wei2024generative}. Unlike the systems in the main text, this benchmark provides a testbed with a significantly higher state-space dimension, allowing for a rigorous evaluation of SEDC's scalability.

\textbf{Experimental Setup}
The objective is to control the opening angle of the jellyfish's wings to achieve a target locomotion pattern. The system state is a high-dimensional PDE field representing the fluid velocity and pressure, while the control input is a scalar time series representing the wing angle. 
\begin{itemize}
    \item \textbf{State Representation:} Each state at a given timestep is represented by a $3 \times 32 \times 32$ tensor, resulting in a \textbf{state dimension of 3,072}. This is a substantial increase in complexity compared to the 1D systems.
    \item \textbf{Dataset:} We generated a dataset of 20,000 control trajectories for training, following the standard procedure for this benchmark.
\end{itemize}

\textbf{Overall Performance Comparison}
We first evaluated SEDC against strong diffusion-based baselines using the full training dataset. The results, shown in Table~\ref{tab:jellyfish_main}, assess control accuracy (Target Loss), energy efficiency (Energy), and computational cost (Training and Inference Time).

\begin{table}[h]
\centering
\caption{Performance comparison on the 2D Jellyfish Locomotion control task (100\% training data). SEDC achieves the best control accuracy with a competitive computational profile.}
\label{tab:jellyfish_main}
\begin{tabular}{lcccc}
\toprule
\textbf{Model} & \textbf{Target Loss} ($\downarrow$) & \textbf{Energy} ($\downarrow$) & \textbf{Train Time} (hrs) & \textbf{Inference Time} (s) \\
\midrule
DecisionDiffuser & 1.74e-4 {\scriptsize$\pm$ 7.4e-5} & 2.016 {\scriptsize$\pm$ 0.08} & 3.0 & 6.0 \\
AdaptDiffuser  & 1.77e-4 {\scriptsize$\pm$ 1.7e-5} & \textbf{2.001 {\scriptsize$\pm$ 0.71}} & 3.0 & 6.0 \\
DiffPhyCon     & 1.77e-4 {\scriptsize$\pm$ 5.4e-5} & 2.157 {\scriptsize$\pm$ 0.21} & 4.0 & 8.0 \\
\textbf{SEDC (Ours)} & \textbf{1.70e-4 {\scriptsize$\pm$ 1.3e-5}} & 2.015 {\scriptsize$\pm$ 0.43} & 3.0 & 6.5 \\
\bottomrule
\end{tabular}
\end{table}

The results demonstrate that SEDC achieves the best control accuracy (lowest Target Loss) among all methods, confirming that its architectural advantages for sample-efficient learning translate effectively to high-dimensional PDE systems. Furthermore, its computational cost remains on par with the fastest baselines, highlighting its practicality.

\textbf{Sample Efficiency Analysis}
A key claim of our work is superior sample efficiency. To specifically validate this on a high-dimensional task, we compared the performance of SEDC against the strongest baseline (DecisionDiffuser) when trained on only 20\% of the available data versus the full dataset.

\begin{table}[h]
\centering
\caption{Sample efficiency comparison on the Jellyfish task. Results show the final Target Loss. SEDC trained on only 20\% of the data outperforms the baseline trained on 100\% of the data.}
\label{tab:jellyfish_efficiency}
\begin{tabular}{lcc}
\toprule
\textbf{Model} & \textbf{100\% Training Data} & \textbf{20\% Training Data} \\
\midrule
DecisionDiffuser & 1.74e-4 & 4.52e-4 \\
\textbf{SEDC (Ours)} & \textbf{1.70e-4} & \textbf{1.75e-4} \\
\bottomrule
\end{tabular}
\end{table}

As shown in Table~\ref{tab:jellyfish_efficiency}, SEDC exhibits remarkable sample efficiency. When trained on just 20\% of the data, its performance is statistically on par with the baseline trained on the full dataset (1.75e-4 vs 1.74e-4). In contrast, the baseline's performance degrades significantly when data is limited. This comprehensive validation on a high-dimensional PDE benchmark strongly supports our paper's central claim: SEDC provides a scalable and highly sample-efficient framework for the control of complex nonlinear systems.

\subsection{System Applicability and Non-Invertible Dynamics Details}
\label{app:non_invertible}

In Section~\ref{sec:case_study}, we demonstrated SEDC's robustness on non-invertible systems. Here we provide the specific dynamics configuration.

\textbf{Rank-Deficient Linear System Setup.}
We constructed a linear system with a rank-deficient control matrix to simulate control redundancy (infinite control solutions):
\begin{equation}
\dot{\mathbf{y}} = \mathbf{A}\mathbf{y} + \mathbf{B}\mathbf{u}, \quad 
\mathbf{A} = \begin{bmatrix} 0 & 1 \\ -1 & -0.5 \end{bmatrix}, \quad 
\mathbf{B} = \begin{bmatrix} 0 & 0 \\ 1 & 1 \end{bmatrix}
\end{equation}
where $\mathbf{y} \in \mathbb{R}^2$ and $\mathbf{u} \in \mathbb{R}^2$. The matrix $\mathbf{B}$ has rank 1, meaning the controls $u_1$ and $u_2$ are coupled ($u_1+u_2$ affects the system), making individual control recovery mathematically impossible.
Despite this, as reported in the main text, SEDC successfully learns a valid control law (implicitly learning an equal distribution or minimum norm solution), whereas the joint diffusion baseline fails to converge to a consistent state-control pair.

\textbf{Additional Experiment: Non-Affine MIMO System.}
We also verified performance on a nonlinear non-affine system: $\dot{x}_1 = -x_1 + x_2$, $\dot{x}_2 = \sin(x_1) - 0.5x_2 + u_1^2 - u_2^2$. SEDC similarly achieved a Target Loss of \textbf{5.0e-2} compared to \textbf{8.1e-2} for the baseline, further confirming robustness.

\subsection{The effectiveness of dynamical learning}
\label{app:consistency}
\begin{figure}[h]
    \centering
    \includegraphics[width=0.7\linewidth]{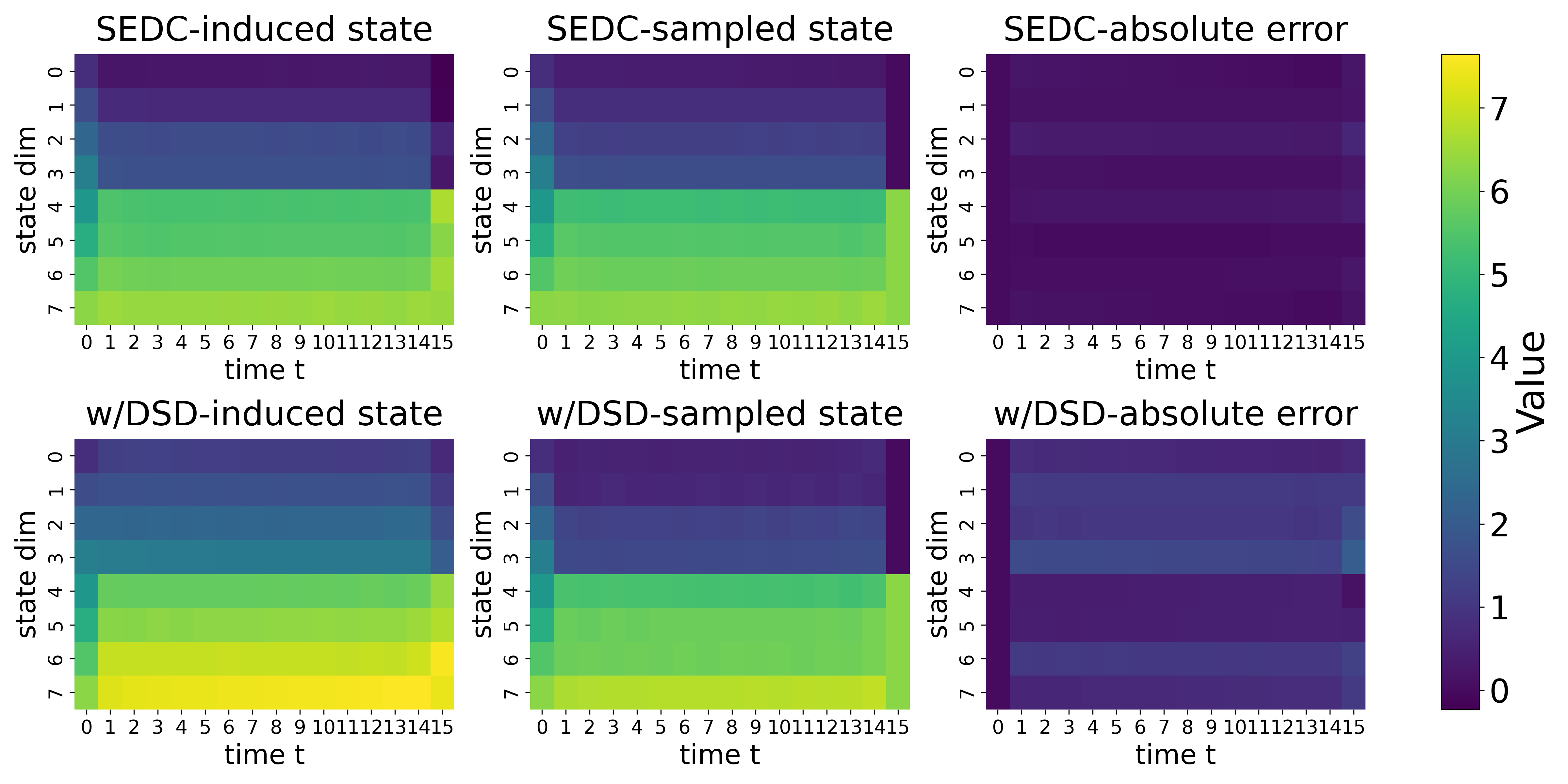}
    \caption{Comparison of State Trajectory Consistency between SEDC and \textit{w/o DSD} Models. The heatmaps show induced states (left), sampled states (middle), and their absolute differences (right) for both SEDC (top) and \textit{w/o DSD} (bottom) approaches under identical start-target conditions.}
    \label{fig:dsd_heatmap}
\end{figure}
To investigate the effectiveness of dynamical learning, we compared the consistency between action sequences and diffusion-sampled state trajectories in models with and without DSD. While both approaches can sample state trajectories from diffusion samples, they differ in action generation: SEDC uses inverse dynamics prediction, whereas \textit{w/o DSD} obtains control signals directly from diffusion samples by simultaneously diffusing states and control inputs. We test both models using identical start-target conditions and visualize the state induced from the generated inputs and the state sampled from the diffusion model, along with the difference~(error) between the above two states in Figure~\ref{fig:dsd_heatmap}. We can observe that SEDC's action-induced state trajectories showed significantly higher consistency with sampled trajectories compared to \textit{w/o DSD}, demonstrating that DSD using inverse dynamics achieves more accurate learning of control-state dynamical relationships.

\subsection{Sufficiency of the 2nd-Order DMD Architecture}
\label{app:dmd_sufficiency}
To address whether our 2nd-order DMD approximation is sufficient for systems with richer nonlinearities, we conducted a targeted experiment on synthetic 2D systems with controlled nonlinear terms. We designed three systems: one purely linear, one with a quadratic term, and one with a cubic term. We compared our full DMD model against a single-UNet baseline and a linear-only ablation of our model.

\begin{table}[h]
\centering
\caption{Performance (Target Loss) on synthetic systems with varying orders of nonlinearity. Our 2nd-order DMD model is sufficient to effectively control the 3rd-order system.}
\label{tab:dmd_sufficiency}
\begin{tabular}{lccc}
\toprule
\textbf{System Dynamics} & \textbf{Single-UNet} & \textbf{DMD (Ours)} & \textbf{Linear-Only} \\
\midrule
1st-Order (Linear) & 1.05e-6 & \textbf{8.68e-7} & 1.54e-6 \\
2nd-Order (Quadratic) & 6.63e-6 & \textbf{5.20e-6} & 2.50e-4 \\
3rd-Order (Cubic) & 7.00e-5 & \textbf{6.43e-5} & 5.80e-3 \\
\bottomrule
\end{tabular}
\end{table}

As shown in Table~\ref{tab:dmd_sufficiency}, our DMD model achieves the lowest error on the 3rd-order system, demonstrating its sufficiency. The performance of the Linear-Only model is nearly 90x worse, confirming that the nonlinear branch ($\mathbf{O}_2$) is not redundant and is critical for capturing the system's dynamics. This principled, sample-efficient design proves robust even for complex systems beyond its explicit Taylor-expansion motivation.

\subsection{Robustness to Observation Noise}
\label{app:noise}
To evaluate the robustness of our method, a critical factor for real-world applicability, we conducted new experiments to validate SEDC's performance when trained on data corrupted by observation noise. This setup simulates practical scenarios where state measurements are imperfect.

\textbf{Experimental Setup}
We added zero-mean Gaussian noise with varying standard deviations ($\sigma$) to the state observations in the training data for both the Kuramoto and Burgers systems. We then retrained our model and the strongest baselines from scratch on this noisy data and evaluated their control accuracy on a clean, noise-free test set.

\textbf{Results}
The results, summarized in Table~\ref{tab:noise_results}, show that SEDC consistently achieves superior control accuracy across all noise levels.

\begin{table}[h]
\centering
\caption{Performance (Target Loss) comparison on noisy training data. Energy cost is shown in parentheses. SEDC demonstrates consistently higher accuracy under noisy conditions.}
\label{tab:noise_results}
\begin{tabular}{l l c c c}
\toprule
\textbf{System} & \textbf{Noise ($\sigma$)} & \textbf{SEDC (Ours)} & \textbf{AdaptDiffuser} & \textbf{DiffPhyCon} \\
\midrule
\multirow{4}{*}{Kuramoto} & 0 & \textbf{8.90e-6} (14.90) & 1.76e-5 (26.23) & 4.80e-4 (18.72) \\
& 0.001 & \textbf{1.21e-5} (17.19) & 8.00e-4 (16.24) & 3.95e-3 (11.50) \\
& 0.01 & \textbf{5.61e-4} (10.40) & 1.75e-3 (3.37) & 3.24e-3 (17.58) \\
& 0.1 & \textbf{3.14e-3} (0.42) & 3.96e-3 (0.98) & \textbf{3.14e-3} (2.22) \\
\midrule
\multirow{4}{*}{Burgers} & 0 & \textbf{9.80e-5} (5.01) & 2.28e-4 (4.65) & 1.62e-4 (5.15) \\
& 0.001 & \textbf{1.86e-4} (5.01) & 2.81e-4 (4.73) & 6.05e-4 (5.11) \\
& 0.01 & \textbf{2.25e-4} (4.40) & 3.98e-4 (4.21) & 8.69e-4 (4.58) \\
& 0.1 & \textbf{6.97e-4} (4.25) & 1.22e-3 (3.75) & 2.89e-3 (3.13) \\
\bottomrule
\end{tabular}
\end{table}

Notably, on the Burgers system with medium noise ($\sigma = 0.01$), SEDC's target loss of 2.25e-4 is 1.8x better than AdaptDiffuser's and 3.9x better than DiffPhyCon's. This superior robustness stems from our key architectural innovations. DMD's decomposition helps capture the core system dynamics resiliently, while DSD's focus on learning a smoother state-only distribution prevents overfitting to noise, a common issue when modeling complex joint state-control distributions.

\section{Visualization}
\label{sup:visual}
\begin{figure}[h]
    \centering
    \begin{subfigure}[b]{0.3\textwidth}
        \includegraphics[width=\textwidth]{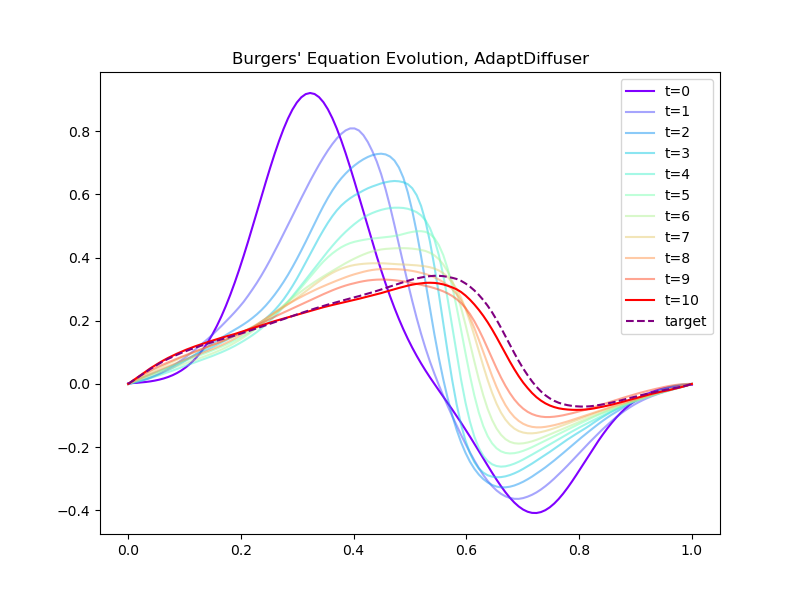}
        \includegraphics[width=\textwidth]{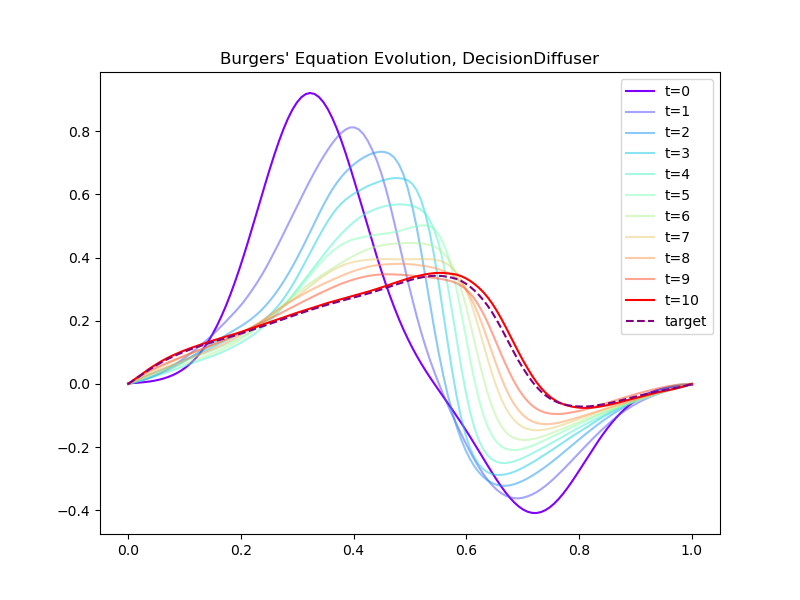}
        \includegraphics[width=\textwidth]{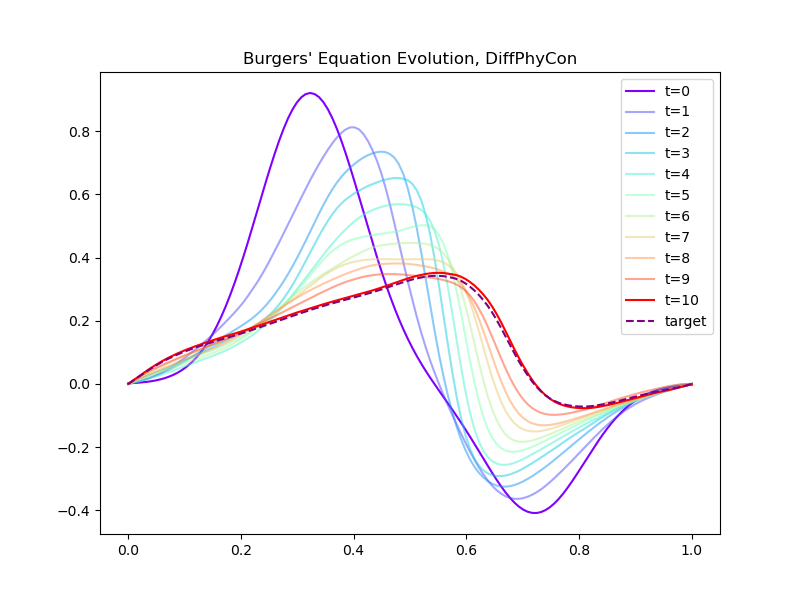}
        \includegraphics[width=\textwidth]{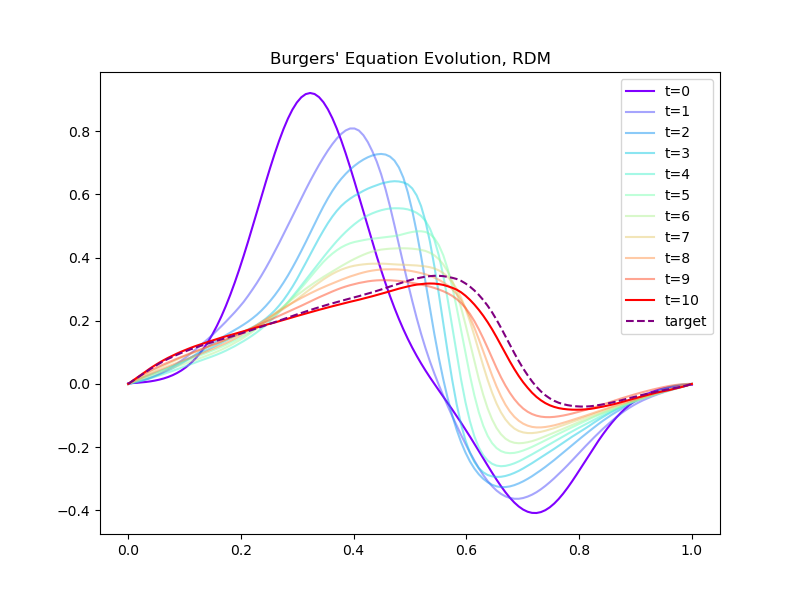}
        \includegraphics[width=\textwidth]{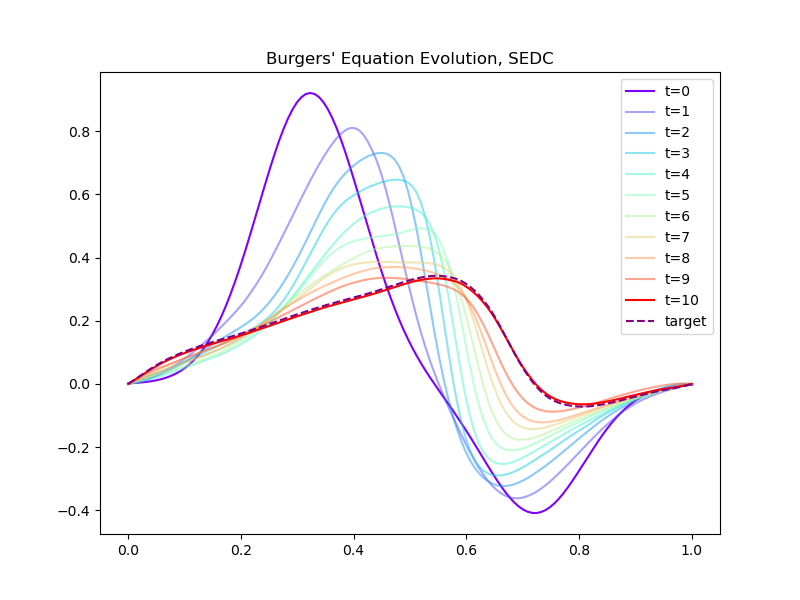}
        \caption{Burgers}
    \end{subfigure}
    \hfill
    \begin{subfigure}[b]{0.3\textwidth}
        \includegraphics[width=\textwidth]{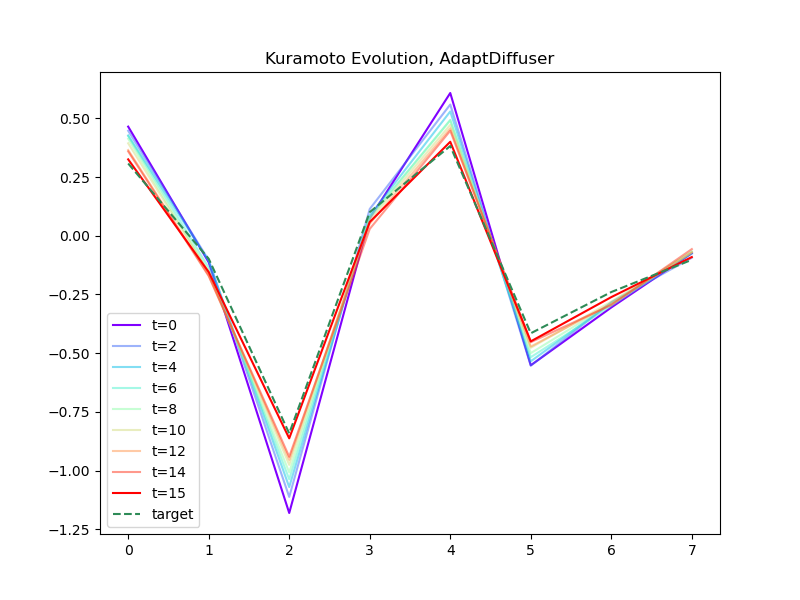}
        \includegraphics[width=\textwidth]{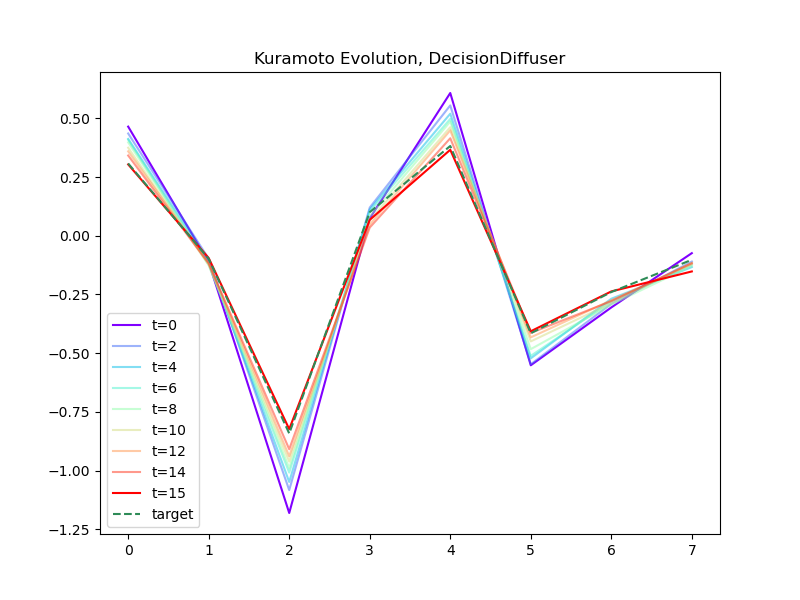}
        \includegraphics[width=\textwidth]{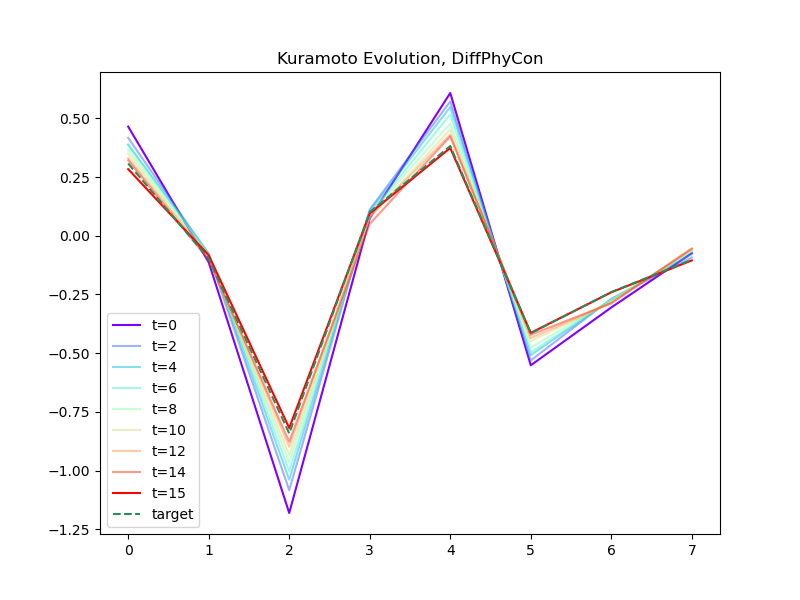}
        \includegraphics[width=\textwidth]{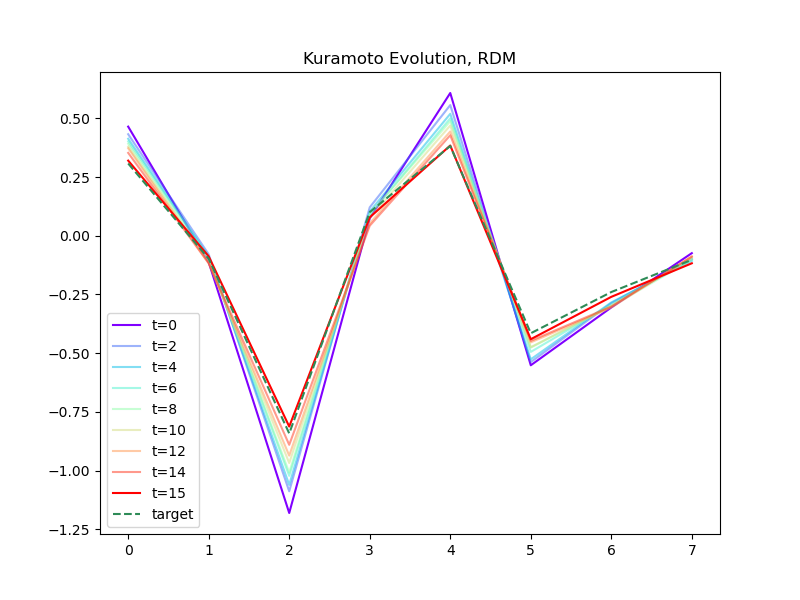}
        \includegraphics[width=\textwidth]{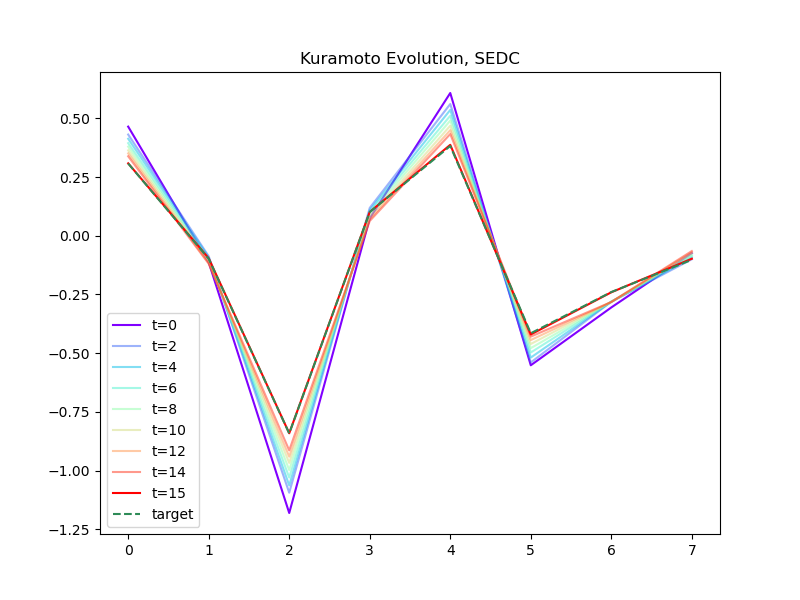}
        \caption{Kuramoto}
    \end{subfigure}
    \hfill
    \begin{subfigure}[b]{0.3\textwidth}
        \includegraphics[width=\textwidth]{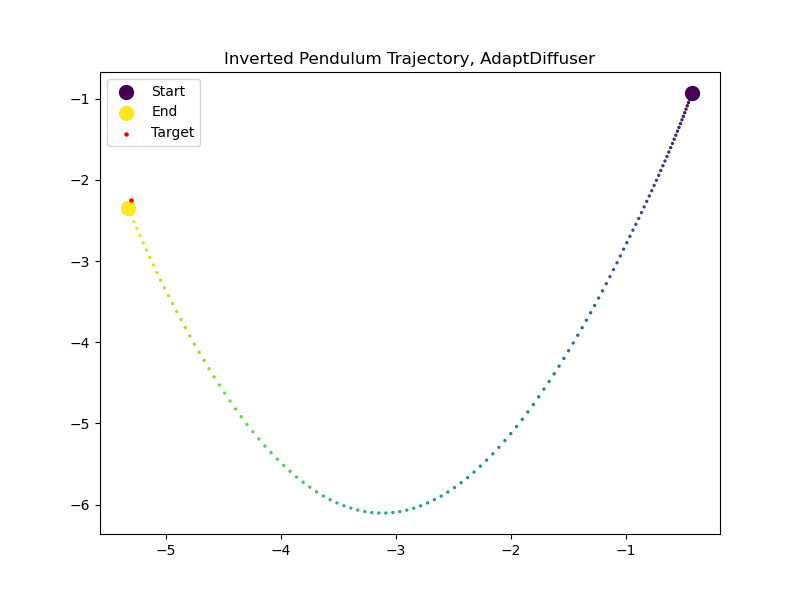}
        \includegraphics[width=\textwidth]{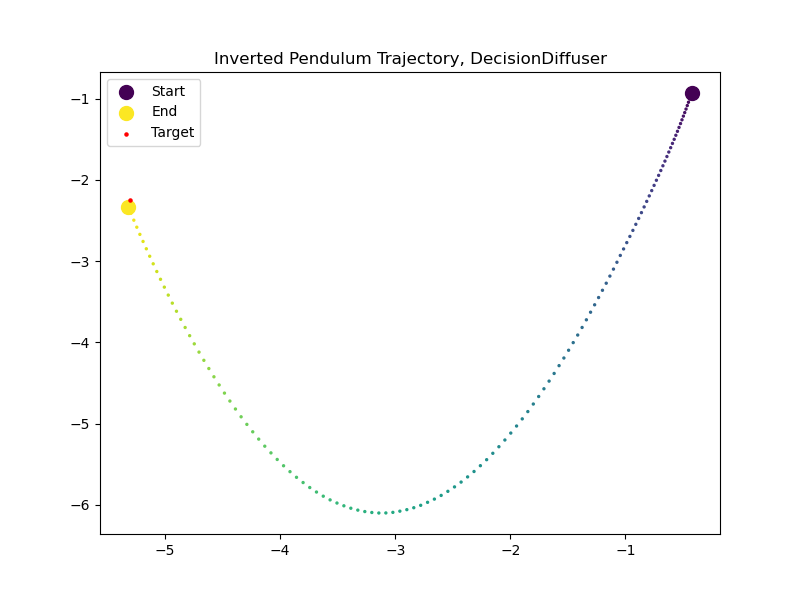}
        \includegraphics[width=\textwidth]{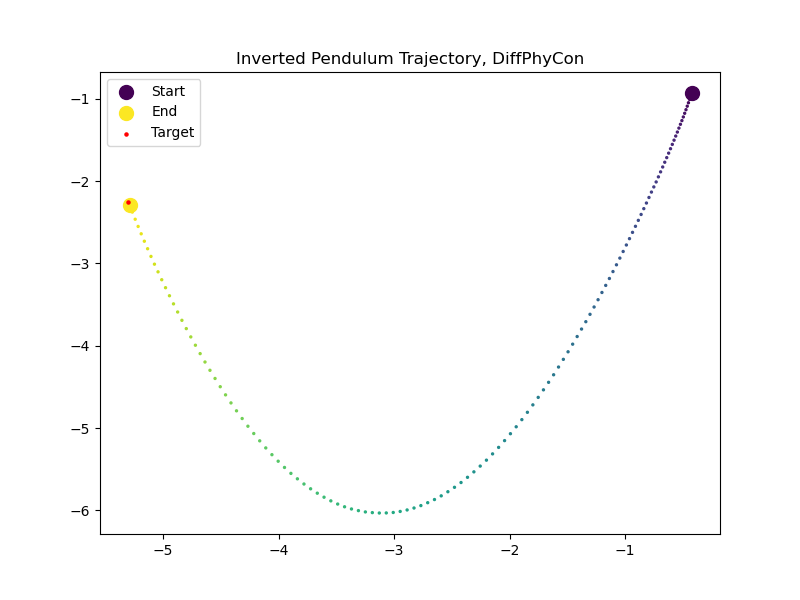}
        \includegraphics[width=\textwidth]{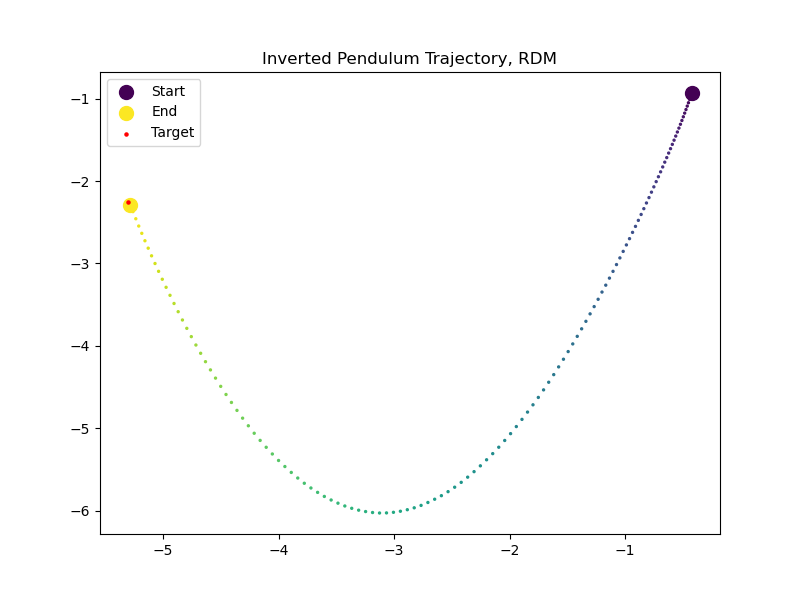}
        \includegraphics[width=\textwidth]{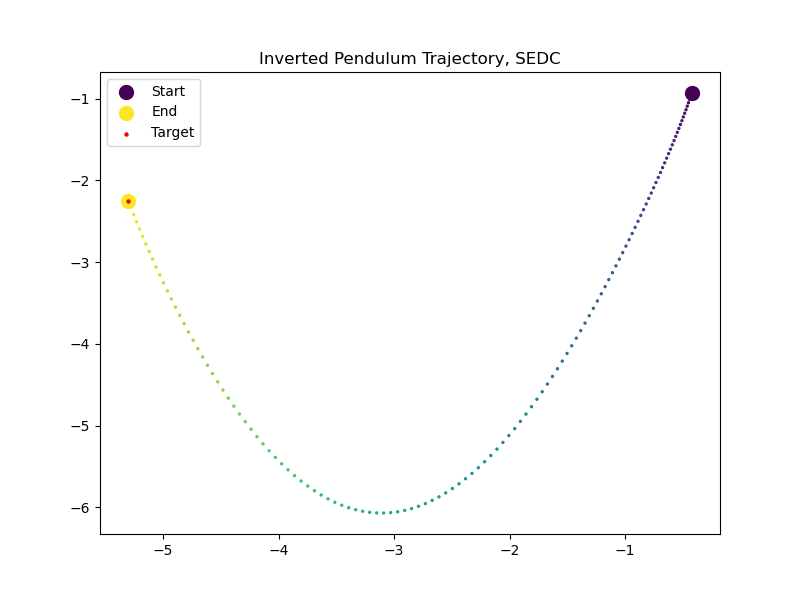}
        \caption{Inverse Pendulum}
    \end{subfigure}
    \caption{Comparison of different methods on Burgers, Kuramoto and Inverse Pendulum systems}
    \label{fig:comparison}
\end{figure}

We present some visualization results of our method and best-performing baselines under three systems. The goal is to make the end state (T=10 for Burgers and T=15 for Kuramoto) close to the target state. As can be seen, SEDC's final state always coincides with the target state. In contrast, the baselines showed inferior results, as some mismatch with the target state can be observed.

\section{Limitations}
In our paper, we assume full state observability throughout. We hope to extend our framework to partial-observable or partial-controllable circumstances in the future.

Considering extending this work to stochastic control settings, our present work focuses on deterministic non-linear systems where SEDC demonstrates significant advantages in sample efficiency and control accuracy. Although we believe the diffusion-based nature of our approach provides a conceptual foundation that could potentially be adapted to stochastic settings, this would require substantial theoretical modifications to our framework components (DSD, DMD, and GSF). Extending to stochastic control would involve addressing additional complexities in modeling state transition probabilities and optimizing over distributions rather than deterministic trajectories. This remains an open research question we are interested in exploring. 

While our current framework generates open-loop trajectories, its computational efficiency and ability to produce high-quality plans make it an ideal candidate for integration into a closed-loop MPC scheme. Such an integration would leverage SEDC as a powerful trajectory planner at each step, enabling robust adaptation to unexpected disturbances by replanning in real-time. This synergy between global planning and reactive feedback represents a promising avenue for future research in robust, sample-efficient, data-driven control.


\end{document}